\documentclass[12pt]{iopart}
\usepackage{iopams}
\usepackage{epsfig} 
\usepackage{graphicx}

\def\gap{\mathrel{\mathpalette\fun >}}

\def\fun#1#2{\lower3.6pt\vbox{\baselineskip0pt\lineskip.9pt
\ialign{$\mathsurround=0pt#1\hfil##\hfil$\crcr#2\crcr\sim\crcr}}}

\def\gap{\mathrel{\mathpalette\fun >}}

\begin{document}

\title[Minimally-parametric power spectrum reconstruction and the evidence for a red tilt]{On Minimally-Parametric Primordial Power Spectrum Reconstruction and the Evidence for a Red Tilt}
\author{Licia Verde$^{1,2}$, Hiranya Peiris$^{3,4}$\\ 
$^1${\it  ICREA \& Institute of Space Sciences (IEEC-CSIC), 
Fac. Ciencies, Campus UAB, Torre C5 parell 2, Bellaterra, Spain}  \\ 
$^2$ {\it Dept. of Astrophysical Sciences, Peyton Hall, Princeton University, Princeton, NJ 08544-1001} \\
$^{3}${\it  Institute of Astronomy, University of Cambridge, Cambridge CB3 0HA, U.K}\\
$^{4}$ {\it Kavli Institute for Cosmological Physics \& Enrico Fermi Institute, University of Chicago, Chicago, IL 60637, U.S.A.} \\
Emails: verde@ieec.uab.es, hiranya@ast.cam.ac.uk}

\begin{abstract}
The latest cosmological data 
seem to indicate a  significant deviation from scale invariance of the primordial power  spectrum when parameterized either by a power law or by a spectral index with non-zero ``running". This deviation, by itself, serves as a powerful tool to discriminate among theories for the origin of cosmological structures such as  inflationary models.  Here, we use a minimally-parametric smoothing spline technique to reconstruct the shape of the primordial power spectrum. This technique is well-suited to search for smooth features in the primordial power spectrum such as deviations from scale invariance or a running spectral index, although it would recover sharp features of high statistical significance. We use the WMAP 3 year results in combination with data from a suite of higher resolution CMB experiments (including the latest ACBAR 2008 release), as well as large-scale structure data from SDSS and 2dFGRS. We employ cross-validation to assess, using the data themselves, the optimal amount of smoothness in the primordial power spectrum consistent with the data.  This minimally-parametric reconstruction supports the  evidence for a power law primordial power spectrum with a red tilt, but not for deviations from a power law power spectrum. Smooth variations in the primordial power spectrum are not significantly degenerate with the other cosmological parameters.

\end{abstract}

\maketitle

\section{Introduction}
\label{sec:intro}

Under simple hypotheses for the shape of the primordial power spectrum, cosmological parameters
 have been measured with exquisite precision from the Wilkinson Microwave Anisotropy Probe (WMAP) data alone \cite{SpergelWMAP03, SpergelWMAP06} or in combination with higher-resolution cosmic microwave background (CMB) experiments  \cite{ACBAR07, CBI04, VSA04,Boom06} and  large scale
structure survey data \cite{SDSS04,2df05}. 

Observations indicate that the primordial power spectrum is consistent with being almost purely adiabatic and close to scale invariant, in agreement with expectations from the simplest inflationary models. Indeed, a power law primordial power spectrum fits both CMB and galaxy survey data very well. Different models for the generation of primordial perturbations yield different deviations from a purely scale invariant spectrum. The simplest can be described in terms of power laws (as e.g., in the simplest slow-roll inflationary models), or a small scale-dependence (``running'') of the spectral index (also in principle arising in inflationary models; see e.g. \cite{Easther:2006tv, Feng:2006ui, Peiris:2006sj} for the implications of the current constraints on this parameter). However, other forms of deviations have been considered: for example, a broken power law \cite{Blanchardetal03,LasenbyDoran05}, an exponential cutoff at large scales \cite{BridgesLasenbyHobson06, BridgesLasenbyHobson07,SpergelWMAP06}, harmonic wiggles superimposed upon a power law arising, for example, from features in the inflaton potential \cite{Starobinsky:1992ts, Adams:2001vc, Okamoto:2003wk, Peiris03, Covietal06, Hamann07, Joy:2007na}, transplanckian physics  \cite{MartinRingeval04, EastherKinneyPeiris05}, multiple inflation \cite{HuntSarkar04}, or ``stringy'' effects in brane inflation models \cite{Bean:2007eh}. 

The statistical significance of such deviations from simple scale invariance is often difficult to interpret \cite{Parkinsonetal06, Pahudetal06, Pahudetal07,Trotta07, GordonTrotta07}. In addition, the significance of these deviations depends on several factors: assumptions about instantaneous  reionization, the treatment of beam uncertanties, treatment of a possible SZ contribution \cite{SpergelWMAP06}, point source subtraction \cite{Huffenbergeretal07, Huffenbergeretal06}, the low  multipole CMB  angular power spectrum ($C_{\ell}$) and likelihood calculation \cite{Eriksenetal07}.

Here, we use  a minimally-parametric reconstruction of the primordial power spectrum  based on that presented in \cite{Sealfon05}. This reconstruction will enable one to answer questions such as: does the signal for the deviation from scale invariance, or deviation from a power-law behavior, come from a localized region in wavelength, or from all scales? In the first case it would be an indication that the assumed functional form is not the correct description of the data. In addition, such an analysis could offer a clue about what could be driving the signal and what systematic effects may most affect the detection. For example, a signal arising only at high multipoles $\ell$ in the WMAP data could point to incorrect noise, beam or point source characterizations. A deviation arising only from the largest observable modes (lowest wavenumber $k$) could, for example, point to foregrounds, the description of low $\ell$ statistics, or assumptions about reionization. As is the case with all non-parametric methods, this approach, not relying on estimates of parameters and their uncertainties, has the drawback that  it cannot provide a  straightforward measurement and a confidence interval. Nevertheless we will discuss how to interpret our results and compare them with parametric methods.
 
Non-parametric or minimally-parametric reconstruction of the primordial power spectrum has only become possible recently, as cosmological datasets now provide enough signal-to-noise to go beyond simple parameter fitting and to explore ``model selection".  While we cannot directly measure the primordial power spectrum, observations of the CMB offer a window into the primordial perturbations  at the largest scales. Large-scale structure (LSS) observations now overlap with scales accessed by CMB observations, and extend the measured $k$ range to smaller scales than the CMB. Both CMB and LSS power spectra depend on the primordial power spectrum via a convolution with a non-linear transfer function which, in turn, depends on the cosmological parameters. In addition, large-scale structure data are affected by galaxy bias and by non-linear  effects, which need to be modeled as outlined below.

Typically, when fitting a model to data, the best fit parameters are found by minimizing the ``distance"\footnote{For example, the chi-square is a distance weighted by the errors.} of the model to the data. But when recovering a continuous function (such as the primordial power spectrum) from discrete data (the CMB $C_{\ell}$ or   the bandpowers for LSS), there are potentially infinite degrees of freedom and finite data points. Thus it is always possible to find at least one function that interpolates the data and has zero or nearly zero ``distance" from the data. However, as the data are noisy, such an interpolation will display features created by the noise (and cosmic variance for cosmological applications) that are not in the true underlying function. On the other hand, if using too few parameters (or the wrong choice of parameterization), the fit could miss real underlying features. Non-parametric or minimally-parametric inference aims at identifying, from the data themselves, how many degrees of freedom are needed to recover the signal without fitting the noise.

Previous work on minimally-parametric reconstruction has employed bins \cite{Bridleetal03,SpergelWMAP06},  piecewise linear reconstruction \cite{Hannestad04}  or a combination \cite{BridgesLasenbyHobson06,BridgesLasenbyHobson07}.
Purely non-parametric techniques involving transfer function deconvolution to directly re-create the power spectrum \cite{ShafielooSouradeep04, ShafielooSouradeep07, Kogoetal04,Tocchini-Valentinietal06, Tocchini-Valentinietal05}, as is the case for all non-parametric methods, may show a  tendency for the recovered function to ``fit the noise''. Wavelets \cite{MukherjeeWang05}  and principal components \cite{Leach05} provide rigorous non-parametric methods to search for sharp features as well as trends in the power spectrum. Some of the techniques presented here, such as cross-validation, could be useful in choosing the number of basis functions to use in these methods.

This paper is organized as follows. First, we build upon the spline reconstruction technique presented in \cite{Sealfon05}, which we briefly review in \S~2. In \S~3 we  apply the method to WMAP third year data (WMAP3) alone and in combination with higher resolution CMB experiments and LSS data. We present our conclusions in \S~4.

\section{Smoothing Spline and Penalized Likelihood}

Since the simplest inflationary models, which are consistent with the data, predict  the primordial power spectrum to be a smooth function, we search for smooth deviations from scale invariance with a cubic smoothing spline technique.  Here we briefly review refs. \cite{GreenSilverman}  and \cite{Sealfon05}. Spline techniques are used to recover a function $f(x)$ based on measurements of $f$ denoted by $\hat{f}$ at $n$ discrete  points $x_i$. Values at $N$  ``knots" of $x$ are chosen.  The values of $F$ (the spline function) at the $N$ knots uniquely define the piecewise cubic spline once we ask for continuity of $F(x)$, its first and second derivative at the knots, and two boundary conditions. We choose to require the second derivative to vanish at the exterior knots.

Allowing infinite freedom to the knot values and simply minimizing the chi-square will tend  in general to fit  features created by the random noise present in the data. It is therefore customary to add a roughness penalty which we chose to be the integral of the second derivative of the spline function\footnote{Note that for a twice continuously differentiable function such as our cubic spline, this is a measure of its total curvature.}
\begin{equation}
\!\!\!\!S(F)=\sum_{i,j=1}^{n}\left[F(x_i)-\hat{f}(x_i)\right]\sigma^{-1}_{ij}\left[F(x_j)-\hat{f}(x_j)\right] + \lambda\int_{x_1}^{x_n}\left[F^{''}(x)\right]^2dx
 \label{eq:penalty}
\end{equation}
where $\sigma_{ij}$ denotes the data covariance matrix, and $\lambda$ is the smoothing parameter. The roughness penalty effectively reduces the degrees of freedom, disfavouring jagged functions that ``fit the noise''. As $\lambda$ goes to infinity, one effectively implements linear regression; as $\lambda$ goes to zero one is interpolating. It can be shown that for this functional form of the penalty function, the cubic spline is the function that minimizes the roughness penalty for given values of $F(x_k)$ at the knots $x_k$. 

The number of knots is usually chosen and  kept fixed in the analysis.  We choose to use $5$--$6$ knots: the dimensionality of the problem grows with the number of knots, thus this corresponds to a $9$-- $10$-- dimensional problem. Beyond a minimum number of knots, there is a trade-off between the number of knots and the penalty, and the form of the  reconstructed function does not depend significantly on the number of knots after this minimum number is reached. As the main goal of this work is to explore, in a minimally parametric way, smooth  deviations from scale invariance (e.g., a red tilt or a running),  a few ($\sim 3$) knots are sufficient. 

In generic applications of smoothing splines, cross-validation is a rigorous statistical technique for choosing the optimal smoothing parameter. Cross-validation (CV) quantifies the notion that if the underlying  function has been correctly recovered, it should accurately predict new, independent data. The most rigorous (but also more  computationally expensive) form of CV is refered to as ``leave-one-out'' CV: the analysis is carried out leaving one data point out, then the distance between the recovered function and that data point is  computed and stored. This is repeated for each data point and then the sum of  the resulting distances, the ``CV score'', is computed. Finally, the best penalty $\lambda$ is the one that minimizes the sum, i.e. the CV score.
 
 $k$-fold cross-validation follows a similar procedure but splits the sample into $k$ subsamples. It becomes identical to ``leave-one-out" CV when $k$ is equal to the number of data points $n$.  For $k\ll n$ this CV technique becomes increasingly faster.  In many cases, for example when the measurements give a direct estimate of $f$ in Eq.~(\ref{eq:penalty}), the calculation of  ``leave-one-out" CV can be significantly shortened \cite{GreenSilverman}. In cosmological applications, however,  the observable quantities  and  the primordial power spectrum are connected by a convolution with the transfer function, making the short-cut not applicable.  
To make the problem computationally manageable, we opt for a $n/2$-fold cross-validation.

 \section{Implementation of Spline Reconstruction}
 
 We consider the following datasets: WMAP three year temperature and polarization power spectra (WMAP3)  \cite{SpergelWMAP06, HinshawWMAP06, PageWMAP06}; Cosmic Background Imager temperature data \cite{CBI04} (CBI); BOOMERanG \cite{Boom06};  Very Small Array temperature power spectrum \cite{VSA04} (VSA);  Arcminute Bolometer Array Receiver temperature power spectrum \cite{ACBAR03, ACBAR07} (ACBAR); \cite{ACBAR08} (ACBAR08); the galaxy power spectrum from the   SDSS  main sample \cite{SDSS04};  the Anglo-Australian Two Degree Field  galaxy redshift survey  \cite{2df05} (2dFGRS) and the power spectrum from the SDSS DR4 luminous red galaxy sample (LRG) \cite{TegmarkLRGDR4}.
 
In particular we consider WMAP3 data alone, and in combination with either higher resolution CMB experiments or large-scale structure data. 

In this application, Eq.~(\ref{eq:penalty}) becomes:
\begin{equation}
\log{\cal L}=\log{\cal L}\left({\rm Data} | {\bf \alpha}, P\left(k\right)\right) + \lambda\int_{k}\left[P^{''}(k)\right]^2d \ln k
\end{equation}
where ${\cal L}\left({\rm Data} | {\bf \alpha}, P\left(k\right)\right)$ denotes the likelihood of the data ($C_{\ell}$ bandpowers, or bandpowers of the  galaxy power spectrum), given the cosmological parameters \{${\bf \alpha}$\} and the primordial power spectrum $P(k)$. In this approach $P(k)$ is fully determined by  its values at the knots.
In other words, as the function to be reconstructed in a minimally parametric way with the spline approach  is $P(k)$, the penalty function following e.g., \cite{GreenSilverman} should be its second derivative $P''(k)$. Another possibility would have been to  parameterize the primordial power spectrum as $\propto k^{n(k)}$ and to  reconstruct the function $n(k)$; in this case the penalty function would have been different, but an underlying assumption on the form of the primordial power spectrum would have been made. 

\begin{figure}
\includegraphics[width=0.53\textwidth, angle=0]{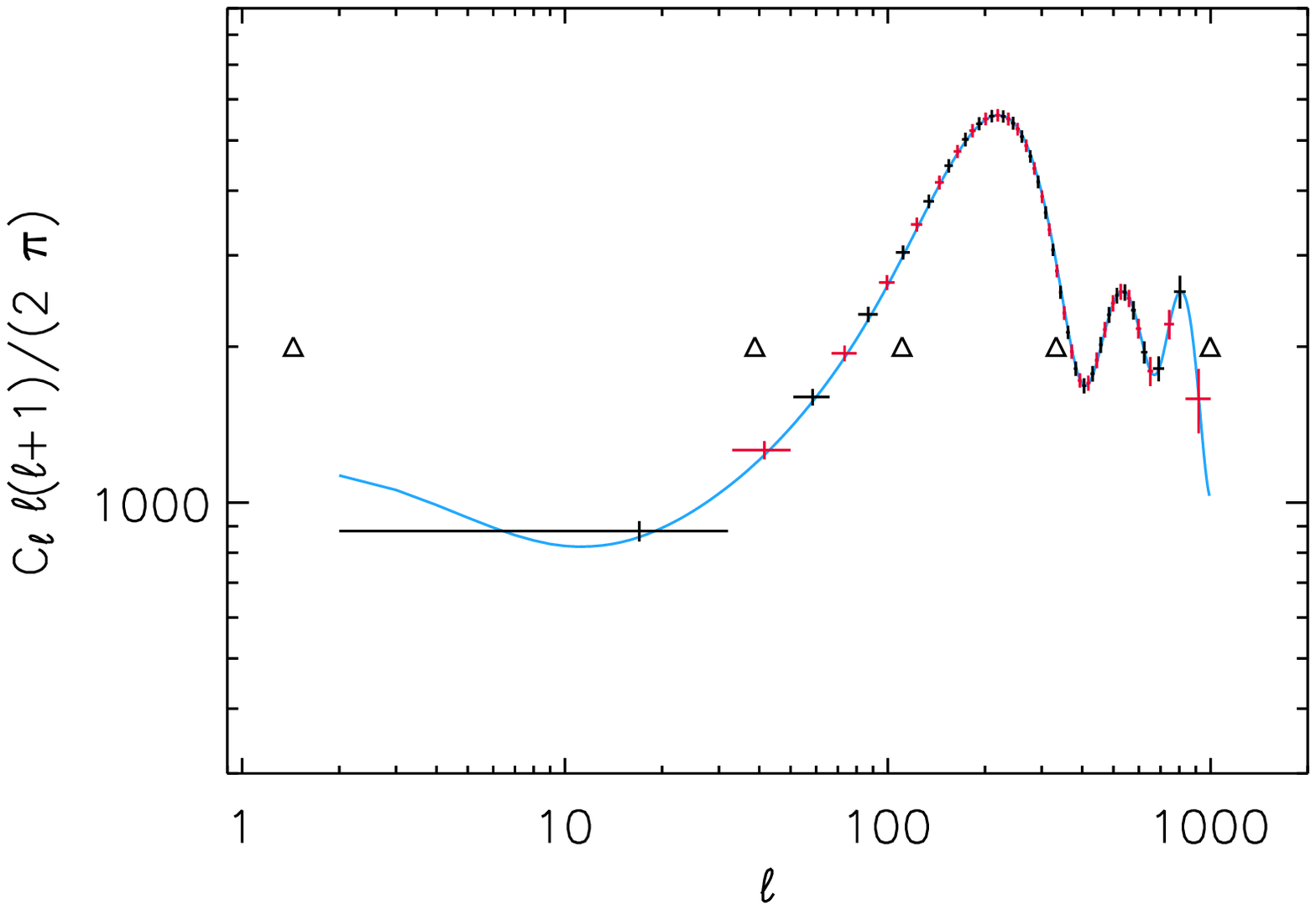}
\includegraphics[width=0.53\textwidth, angle=0]{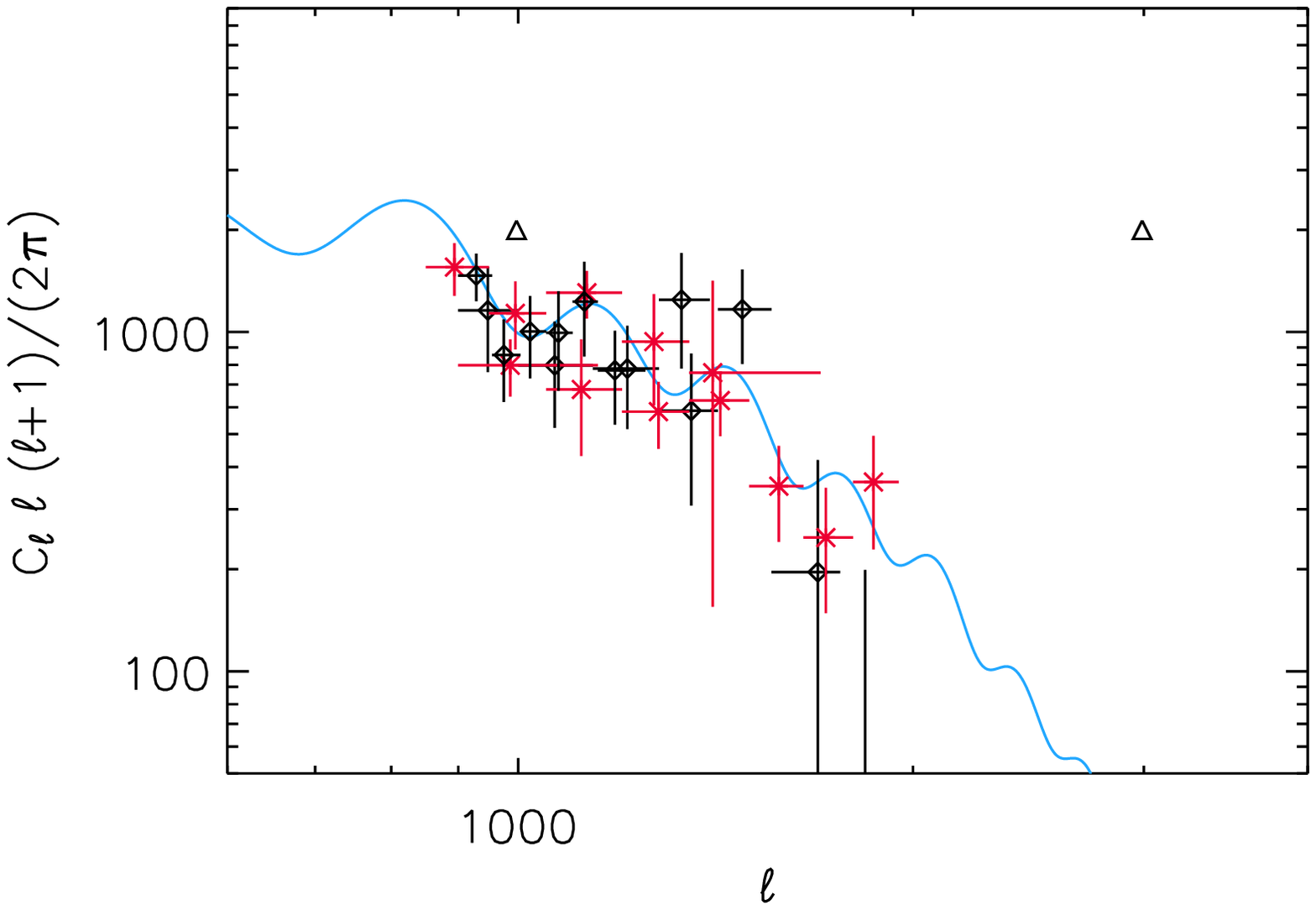}
\caption{Left: Triangles show the position of the 5 knots used for WMAP3. Crosses show the CV bins (CV1 in red , CV2 in black); the width of the cross shows the bin width and the height shows the expected noise (cosmic variance + experimental noise) for the  fiducial model shown by the solid line. Right:  CV set up for higher-resolution CMB experiments, the color scheme is as in the left panel. The points are the  actual data with error-bars. The datasets used are CBI, VSA for CV1 and  BOOMERanG,VSA for CV2. Throughout $C_{\ell}$ are in units of $\mu K^2$. }
\label{fig:knots}
\end{figure}

We use 5 knots for the WMAP3 data when considered alone and 6 knots when in combination with other datasets.  As explained above, our main goal is to explore smooth  deviations from scale invariance and thus a few ($\sim 3$) knots are sufficient.  
The knot locations are illustrated in Fig. \ref{fig:knots}. 
We have explored different knot locations and found that while the reconstructed form for $P(k)$ does not depend significantly on knot locations (as long as the knots sample the full $k$-range, see Appendix) the convergence speed of the Markov chains does depend on knot location.

We further develop  the implementation of \cite{Sealfon05} by (a) running new chains for each value of the penalty rather than using importance sampling to explore small changes in $\lambda$; (b) varying cosmological parameters  when computing CV; (c) adding other datasets beyond WMAP3 to the data compilation; and (d) changing the way the CV sample is chosen.

To set up CV, the data set is split into two samples (denoted by CV1 and CV2). We split WMAP3 data in bins of  roughly equal signal-to-noise, as illustrated in Fig. \ref{fig:knots}. In the released WMAP3 v2p2p2 likelihood package, the low $\ell$ likelihood ($\ell < 32$) is computed using a pixel-based method, and thus $\ell$'s from 2 to 32 must belong to the same CV bin.  This sets the CV bin size: all the CV bins have roughly the same  signal-to-noise. With this choice we also minimize the effect of off-diagonal coupling (which becomes negligible at large separations in $\ell$). The polarization data (TE and EE) at low $\ell < 23$ is always used in implementing CV, as it encodes information on reionization and the optical depth parameter $\tau$, and not on the shape of the primordial power spectrum. We find that this choice greatly reduces the degeneracy between $\tau$ and the  shape of the primordial power spectrum on the largest scales in the CV1 runs.
Note that for the $k$-range corresponding to  $\ell<100$, there are $2$ knots and $5$ CV bins, while in the range corresponding to $\ell> 100$, there are $3$ knots and $53$ CV bins. As each bin has roughly the same signal-to-noise, the low $\ell$ range is actually sampled by the knots much more finely than the high $\ell$ range.

For the remaining datasets we set up CV as follows. For the high resolution CMB data, CV1 includes  VSA and ACBAR, and CV2 includes CBI and BOOMERanG (see Fig. \ref{fig:knots}). As SDSS bandpowers are essentially uncorrelated, we use every other bandpower for CV.

For each of the CV samples and for a grid of penalty values $\lambda$, we run a Markov Chain Monte Carlo (MCMC), using a suitably modified version of the publicly available software CosmoMC \cite{camb, cosmomc}. The best fit model  from CV1 is then run through the CV2 data sample, the likelihood is stored, and vice versa. For each value of the penalty  $\lambda$, the sum of the logarithm of the two likelihoods so obtained is our proxy for the CV score. The optimal penalty is the one that maximizes the CV score. Once the optimal penalty is found, a MCMC is run for the chosen penalty on using all the data. 

\subsection{WMAP 3-year data  alone}

We start by considering WMAP3 data alone  and use the latest version of the WMAP likelihood code (v2p2p2) which includes an updated point-source correction \cite{Huffenbergeretal06}, beam error propagation  and foreground marginalization on large scales. We do not include a Sunyaev-Zel'dovich \cite{SZ} (SZ) contribution to the $C_{\ell}$. 

\begin{figure}
\includegraphics[width=0.53\textwidth]{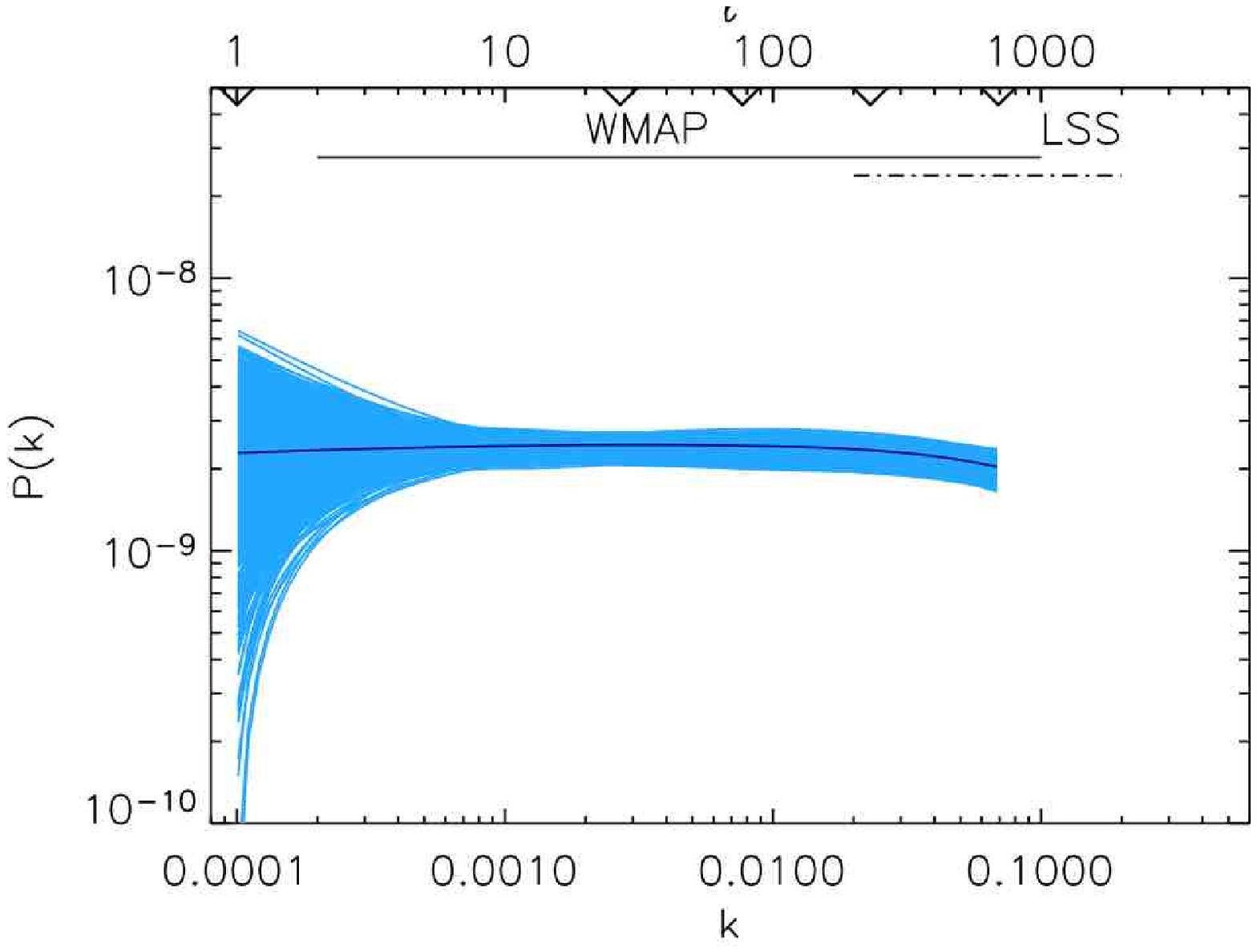}
\includegraphics[width=0.53\textwidth]{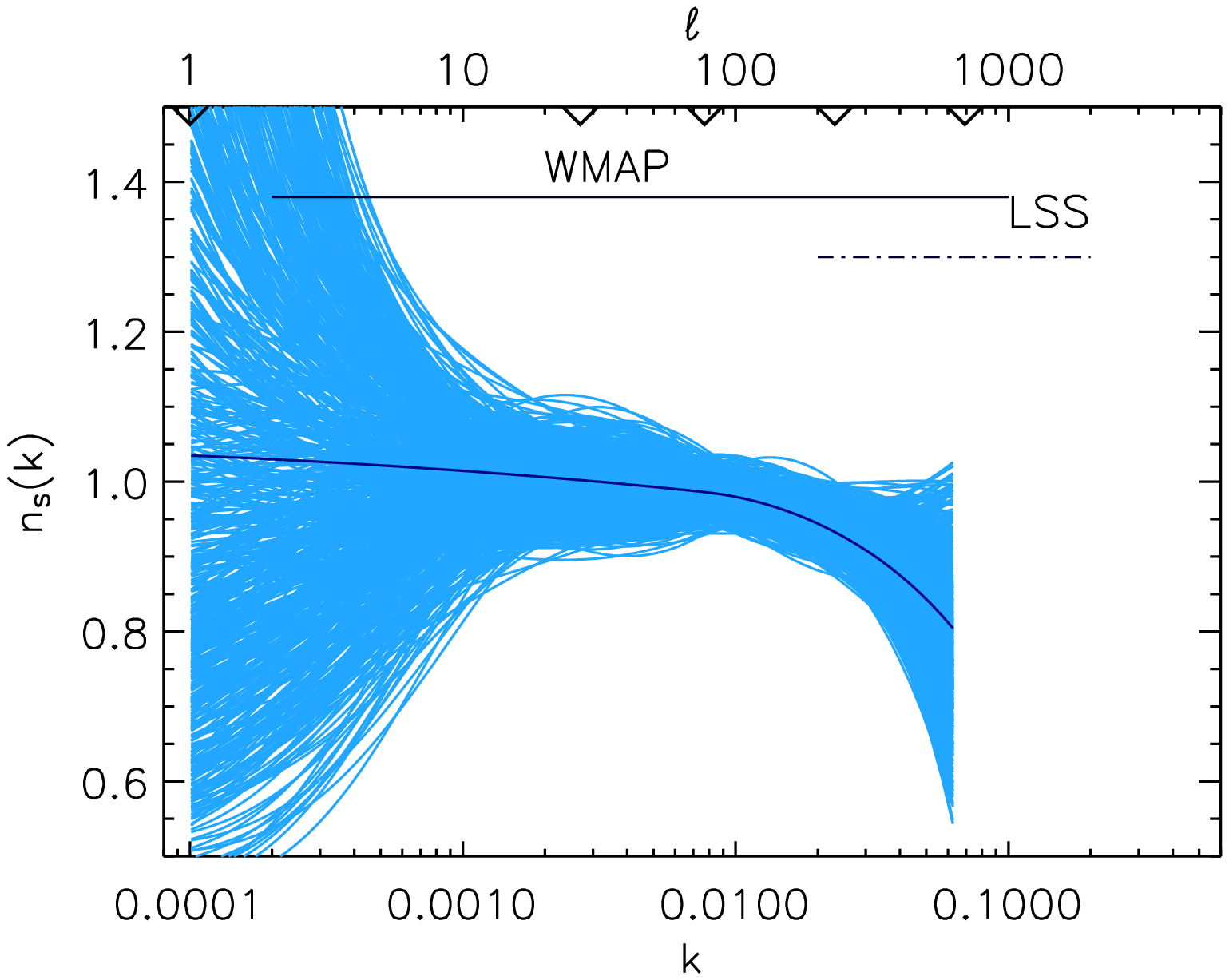}
\caption{Primordial power spectrum $P(k)$ (left) and corresponding spectral index $n_s(k)$ (right) reconstructed from WMAP3 data alone for the CV-selected optimal penalty. The primordial power spectrum shape seems to acquire a curvature at $\ell >300$. The location of the knots is shown on the top of the figure. Throughout, the units of $k$ are Mpc$^{-1}$.}
\label{fig:wmappkoptpenalty}
\end{figure}

In Fig. \ref{fig:wmappkoptpenalty} we show the reconstructed power spectrum for the optimal penalty ($\lambda_{\rm opt, WMAP}$) and  the corresponding spectral index $n_s(k)$. Here $n_s(k)$ is defined (and obtained)  by the first derivative of $P(k)$: $n_s(k)=1+d\ln P/d\ln k$.  The light (blue) lines are the best fitting 68\% and the darker line is the multi-dimensional best fit \footnote{In all cases, when plotting  the 95\%  best fitting spline curves, the ``envelope" is only slightly larger but the individual spline curves are more ``wiggly". For clarity, we show only  the 68\% best curves.}.   As customary in CMB studies,  $k$ is in units of Mpc$^{-1}$.   When interpreting the plot of the scale dependence of the spectral slope $n_s(k)$, one should keep in mind that the quantity that was reconstructed, and for which CV was used to find the optimal penalty, is actually the power spectrum; thus  the penalty may be sub-optimal for $n_s(k)$.

The large cosmic variance in the  low  $\ell$ region allows a lot of freedom in the shape of the primordial power spectrum, but  makes any downturn at large scales not significant. Fig. \ref{fig:wmappkoptpenalty} indicates that  the signal for  deviation from scale invariance seems to arise from $\ell \gap400$, where a downturn is visible. The inclusion of an SZ contribution  would make this effect even larger, but is not considered here.  Beyond this trend at high $\ell$, we do not find any evidence for features in the power spectrum. At high $\ell$, different systematics affect the reconstructed $P(k)$: the beam model, beam error propagation and point source subtraction. In particular \cite{Huffenbergeretal07} find that, assuming a power law primordial power spectrum, the recovered spectral slope changes depending on the estimated point source amplitude and its error, but that beam errors have a small effect. They tentatively argue that the point source uncertainty should be increased by 60\% compared to the WMAP estimated value (which would tend to reduce the significance of a red spectrum). They also suggest  that  the fiducial point source contribution to be subtracted out may be smaller  by $\sim 25$\% than the WMAP value. We find that, if we use the fiducial  point source amplitude estimate of  \cite{Huffenbergeretal07}, the reconstructed $P(k)$ does not show the high $\ell$ downturn.  Along the same lines, \cite{ACBAR08}  find that there is a tension between the  $\sigma_8$ value recovered from WMAP3 data alone and that recovered from WMAP3+ ACBAR08 data, with WMAP's estimate being lower.   They conclude that the lower $\sigma_8$ value favored by WMAP3 alone is driven by WMAP measurements  at  high $\ell$.

It is interesting to compare this reconstruction with that presented in \cite{Sealfon05} for the first year WMAP data release (WMAP1). A direct comparison of the two studies needs to be done with caution. They were interested in deviations from scale invariance; thus they report the quantity  $\hat{n}(k)=1+(\ln P(k)-\ln P(k_0))/(\ln k-\ln k_0)$, while here we are interested in more general deviations, so we show $n_s(k)$. For $n_s=1$, $\hat{n}\equiv n$. We can see that WMAP1-based reconstruction shows a similar behavior: a $P(k)$ consistent with scale invariance on large scales and a downturn at $k>0.01$. But the WMAP3 optimal penalty is lower than that for WMAP1 ($0.02$ vs $0.1$ when converted to the same units), reflecting the fact that the noise level in WMAP3 is lower (in particular, the error on $C_{\ell}$ due to noise is a factor $\sim 3$ lower). 

In Table \ref{tab:wmaponly} we report constraints on cosmological parameters to show how they are affected by the extra freedom in the primordial power spectrum, along with the power law and running spectral index models as reported in \cite{SpergelWMAP06}. The $\tau$ determination is virtually unaffected by the additional freedom in the primordial power spectrum. This was not the case in WMAP1 (see \cite{Sealfon05}), but this is understandable as, in WMAP3, $\tau$ is well constrained by the  $EE$ polarization data alone \cite{PageWMAP06}.
 
\begin{table}
\caption{\label{tab:wmaponly} Effect on cosmological parameters of the extra freedom in the primordial power spectrum, for  WMAP3 alone. We report only the parameters for which errors are affected more than 10\%. ``PL" means power law power spectrum,``run" means running spectral index, ``spline $\lambda_{\rm opt, WMAP}$" means spline reconstruction with optimal penalty set by CV, and ``spline $\lambda=0$" means spline with no penalty. ``PL" and ``run" are taken from \cite{SpergelWMAP06}.}
\begin{center}
\footnotesize\rm
\begin{tabular*}{0.9\textwidth}{@{\extracolsep{\fill}}|l|c|c|c|c|}
\hline
WMAP & PL & run & spline $\lambda_{\rm opt, WMAP}$ & spline $\lambda=0$\\
\hline
$\Omega_b h^2$ & $0.0223\pm 0.00073$ & $0.021\pm 0.001$ & $0.021\pm 0.001$ & $0.0192\pm 0.0012$\\
$\Omega_c h^2$ & $0.1054 \pm 0.0078$ & $0.114\pm 0.0098$ & $0.117 \pm 0.011$ & $0.141 \pm 0.018$\\
$h$  &  $0.733 \pm 0.032$ & $0.681\pm 0.042$ & $0.679 \pm 0.047$ & $0.584 \pm 0.058$\\
$\sigma_8$ & $0.761 \pm 0.049$ & $0.77 \pm 0.05$ &  $0.818 \pm 0.052$ & $0.881 \pm 0.051$\\
\hline
\end{tabular*}
\end{center}
\end{table}
\subsection{WMAP 3-year data  and Higher Resolution CMB Experiments}

Following \cite{SpergelWMAP06}, to minimize covariance between WMAP and  higher resolution CMB experiments, we consider only the following subsets of the data:
for ACBAR, only bandpowers at $\ell >800$; for CBI, only bandpowers 5 to  12 ($600< \ell < 1800$); for VSA, 5 band powers with mean $\ell$-values of 894, 995, 1117, 1269  and 1407;  and for BOOMERanG, 7 bandpowers with central  $\ell >800$. As before, we do not consider an SZ contribution to the $C_{\ell}$, but, by not considering band powers at $\ell>2000$,  scales  possibly affected by the ``SZ excess" are excluded from the analysis.

When combining WMAP3 with higher resolution CMB experiments (WMAPext), we find that the optimal penalty ($\lambda_{\rm opt, ext}$) becomes higher ($\lambda_{\rm opt, ext}= 25 \lambda_{\rm opt, WMAP}$; i.e. the data  do not require as much freedom in the shape of the primordial power spectrum), and that CV becomes less sensitive to the value of the penalty. In other words, the CV score dependence on penalty flattens out. We interpret this as the recovered $P(k)$ being smooth, and its second derivative being small enough to make the total likelihood less sensitive to the penalty function. In fact, most  $P(k)$ features giving rise to a second derivative  are localized at  low $\ell$ (small $k$) where the signal is dominated by  cosmic variance. As statistical power is added to small scales,  the ``wiggliness" allowed by the large scales gets downweighted. 
Since the WMAP3 penalty is lower than the one found for WMAP1 by \cite{Sealfon05}, one may intuitively expect that adding extra datasets would reduce the penalty further. Here this is not the case: first, an extra knot is added and the  $k$-range probed  increases by a decade; second,  the downturn that was significant in WMAP3 data alone is now not as significant. 

\begin{figure}
\includegraphics[width=0.53\textwidth]{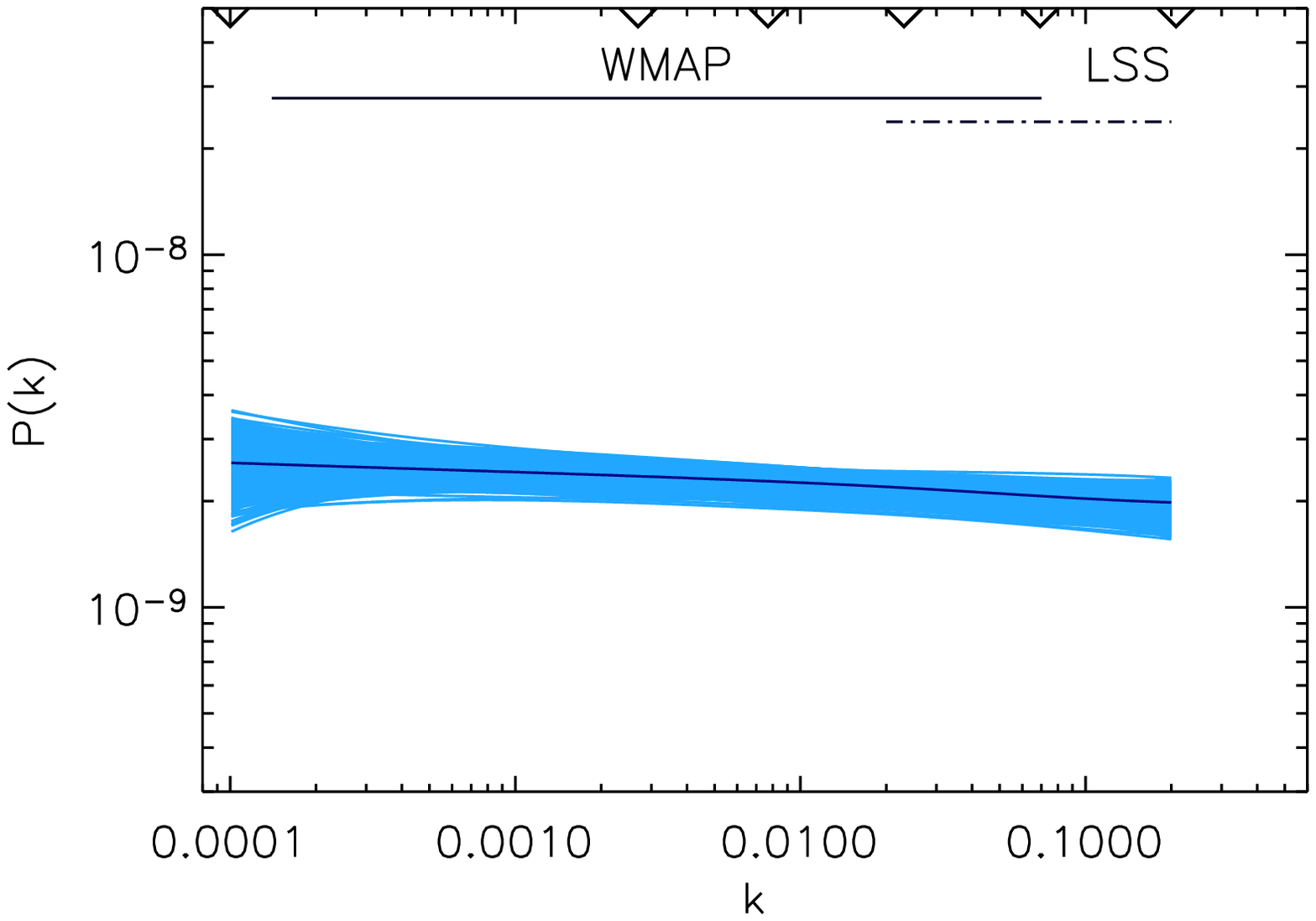}
\includegraphics[width=0.53\textwidth]{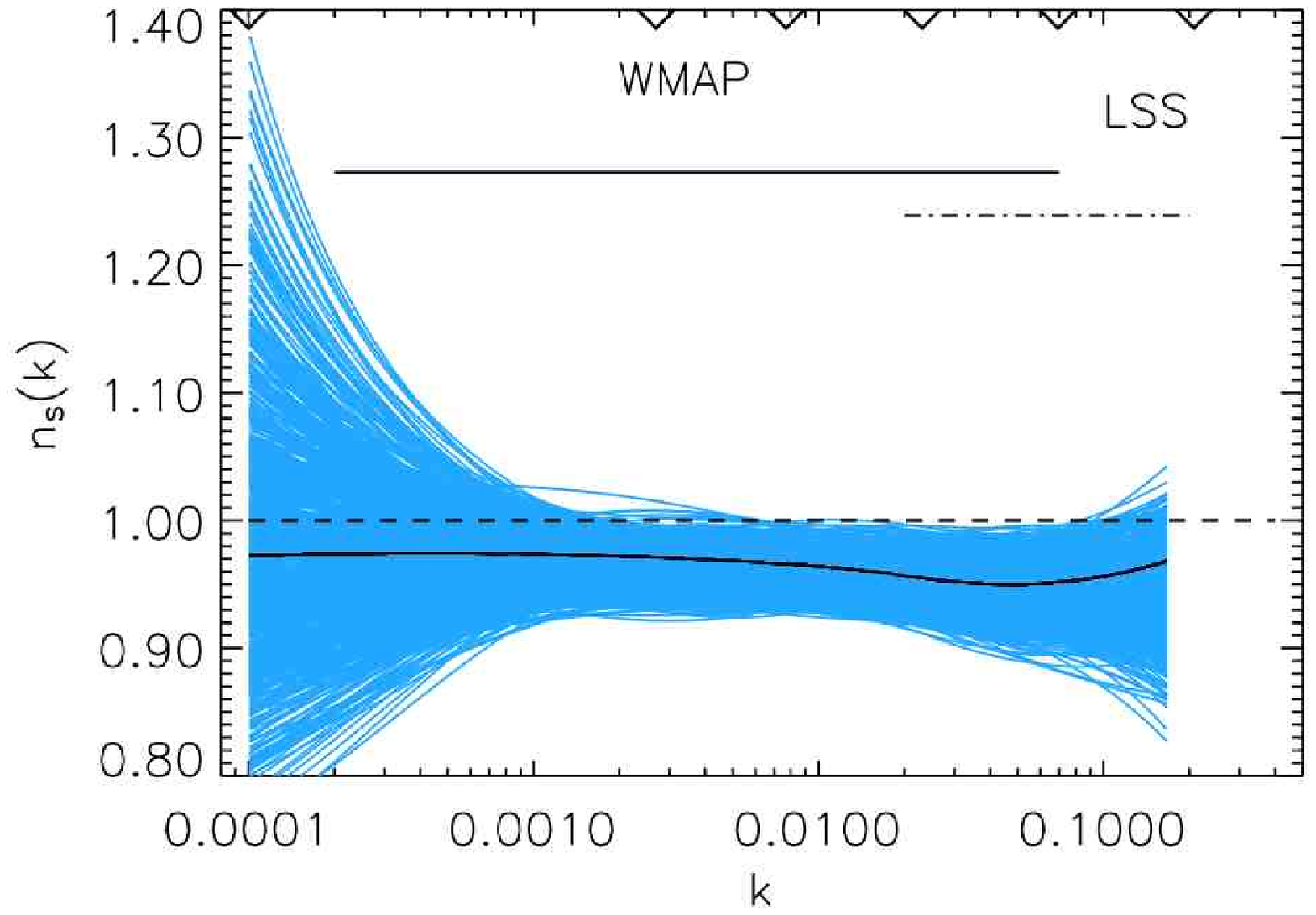}
\caption{Primordial power spectrum $P(k)$ (left) and spectral index $n_s(k)$ (right) reconstructed from WMAPext for the CV-selected optimal penalty. A deviation from scale invariance consistent with a red-tilted power law form is clearly visible. The dashed line corresponds to a scale invariant power spectrum: the reconstructed spectrum is consistent with a scale independent spectral slope and a red tilt. Throughout, the units of $k$ are Mpc$^{-1}$.}
\label{fig:wmapextpkhi5}
\end{figure}

Fig. \ref{fig:wmapextpkhi5} shows the $P(k)$ recovered for the optimal (CV-selected) penalty.
Now a deviation from scale invariance is clearly visible; the signal is distributed on all scales and consistent with a red-tilted power law power spectrum. This is more clearly seen in the  corresponding dependence  on scale of the spectral slope. A scale-independent  spectral slope and a red tilt is a better fit to the data than a scale invariant power spectrum (indicated by the dashed line).
 
\begin{figure}
\includegraphics[width=0.53\textwidth]{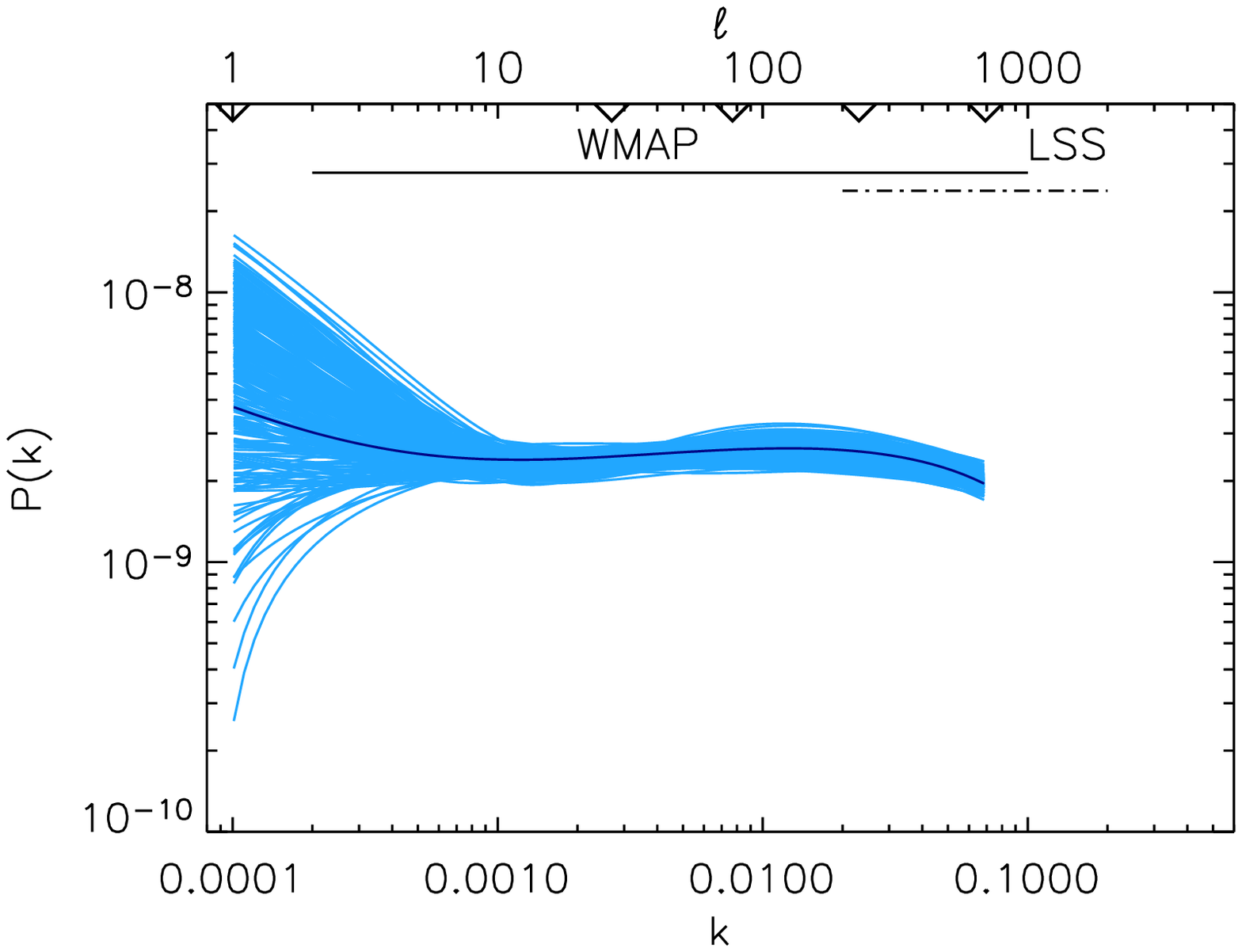}
\includegraphics[width=0.53\textwidth]{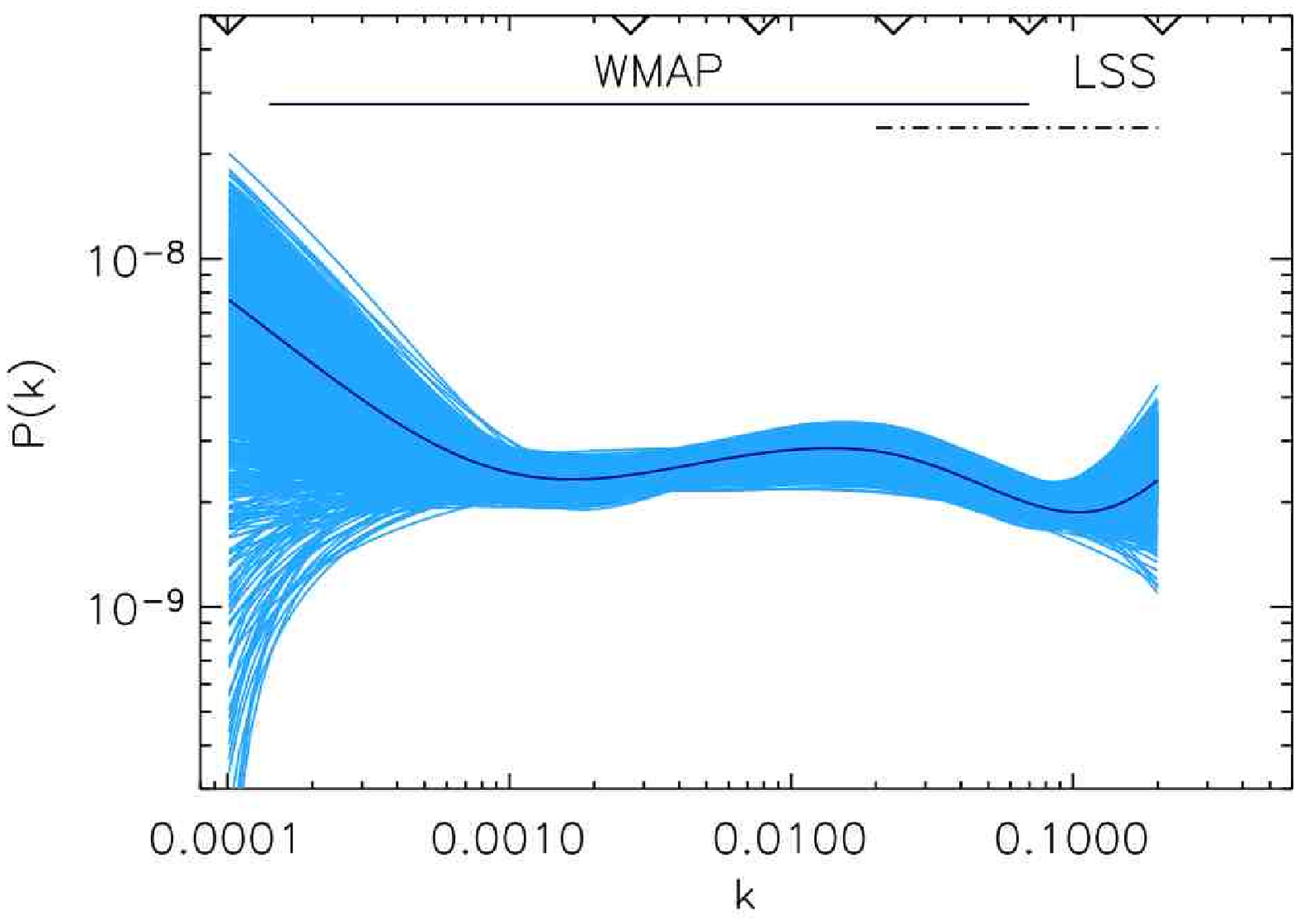}
\caption{Primordial power spectrum $P(k)$ reconstructed from WMAP3 (left) and WMAPext (right) data, without  CV penalty. While one may be tempted to interpret the reconstructed power spectrum as having features, CV shows  that they are not significant, and the recovered optimal $P(k)$ is that shown in Figs. 2 and 3. The units of $k$ are Mpc$^{-1}$.}
\label{fig:pknopenalty}
\end{figure}

For comparison, and to visualize the effect of implementing CV,  in Fig. \ref{fig:pknopenalty} we show the reconstructed power spectrum  for penalty set to zero. While one may be tempted to interpret the reconstructed power spectrum as having features, CV shows  that they are not significant. 
In Table \ref{tab:wmapext} we report the effect on cosmological parameters of the extra freedom in the primordial power spectrum. As we have seen previously, $\tau$ is not affected.
 \begin{table}
\caption{\label{tab:wmapext} Effect on cosmological parameters of the extra freedom in the primordial power spectrum  for WMAPext data, in the same format as Table \ref{tab:wmaponly}.}
\begin{center}
\footnotesize\rm
\begin{tabular*}{0.9\textwidth}{@{\extracolsep{\fill}}|l|c|c|c|c|}
\hline
WMAPext & PL & run & spline $\lambda_{\rm opt, ext}$ & spline $\lambda=0$\\
\hline
$\Omega_b h^2$ & $0.0223\pm 0.00073$ & $0.021\pm 0.001$ & $0.0221\pm 0.00075$ & $0.018\pm 0.0011$\\
$\Omega_c h^2$ & $0.103 \pm 0.0081$ & $0.114\pm 0.0098$ & $0.106\pm 0.0071$ & $0.15 \pm 0.017$\\
$h$  &  $0.739 \pm 0.031$ & $0.68\pm 0.04$ & $0.733 \pm 0.033$ & $0.55 \pm 0.056$\\
$\sigma_8$ & $0.739 \pm 0.049$ & $0.77 \pm 0.05$ &  $0.764 \pm 0.042$ & $0.92 \pm 0.056$\\
\hline
\end{tabular*}
\end{center}
\end{table}

\begin{figure}
\includegraphics[width=0.53\textwidth]{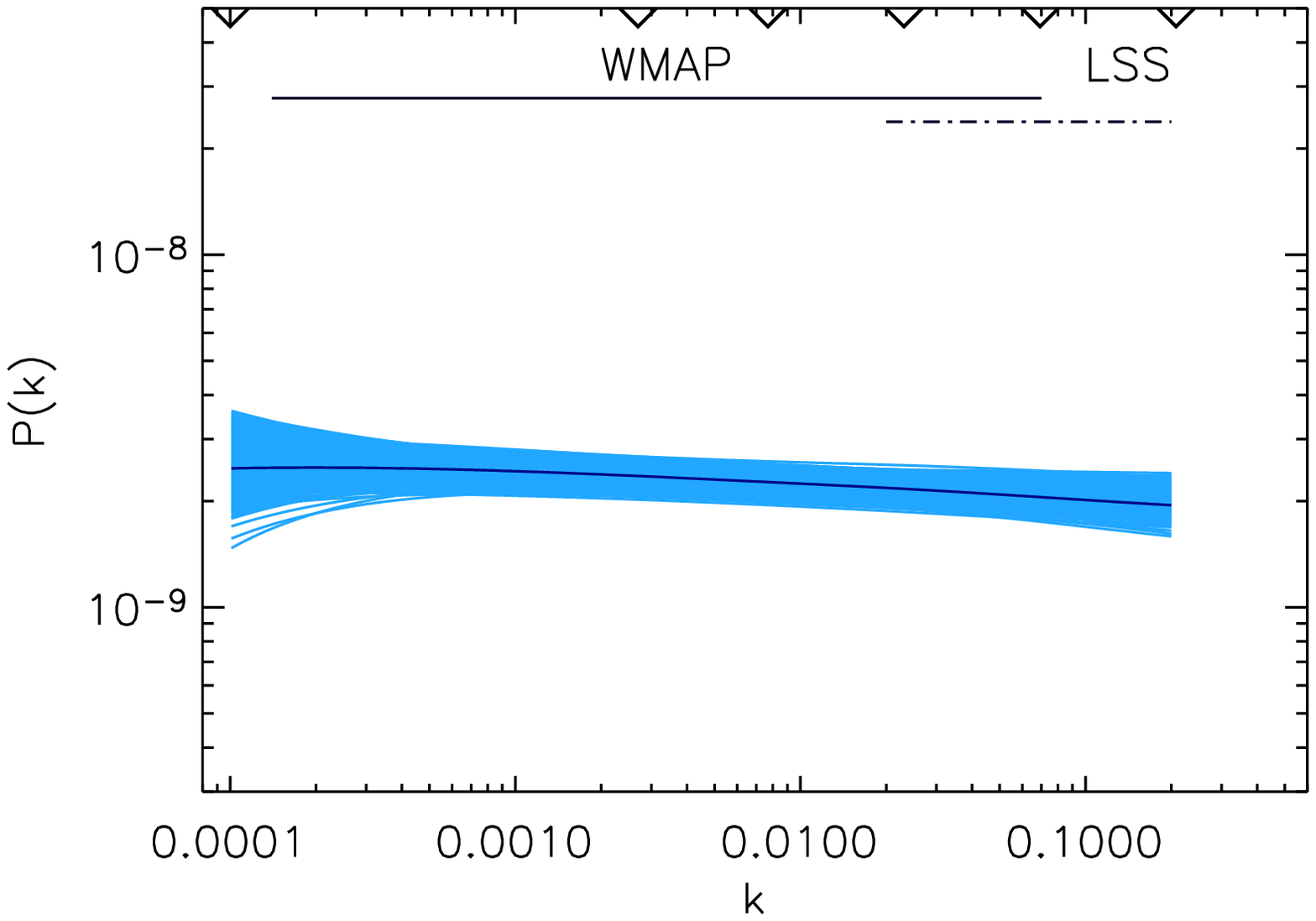}
\includegraphics[width=0.53\textwidth]{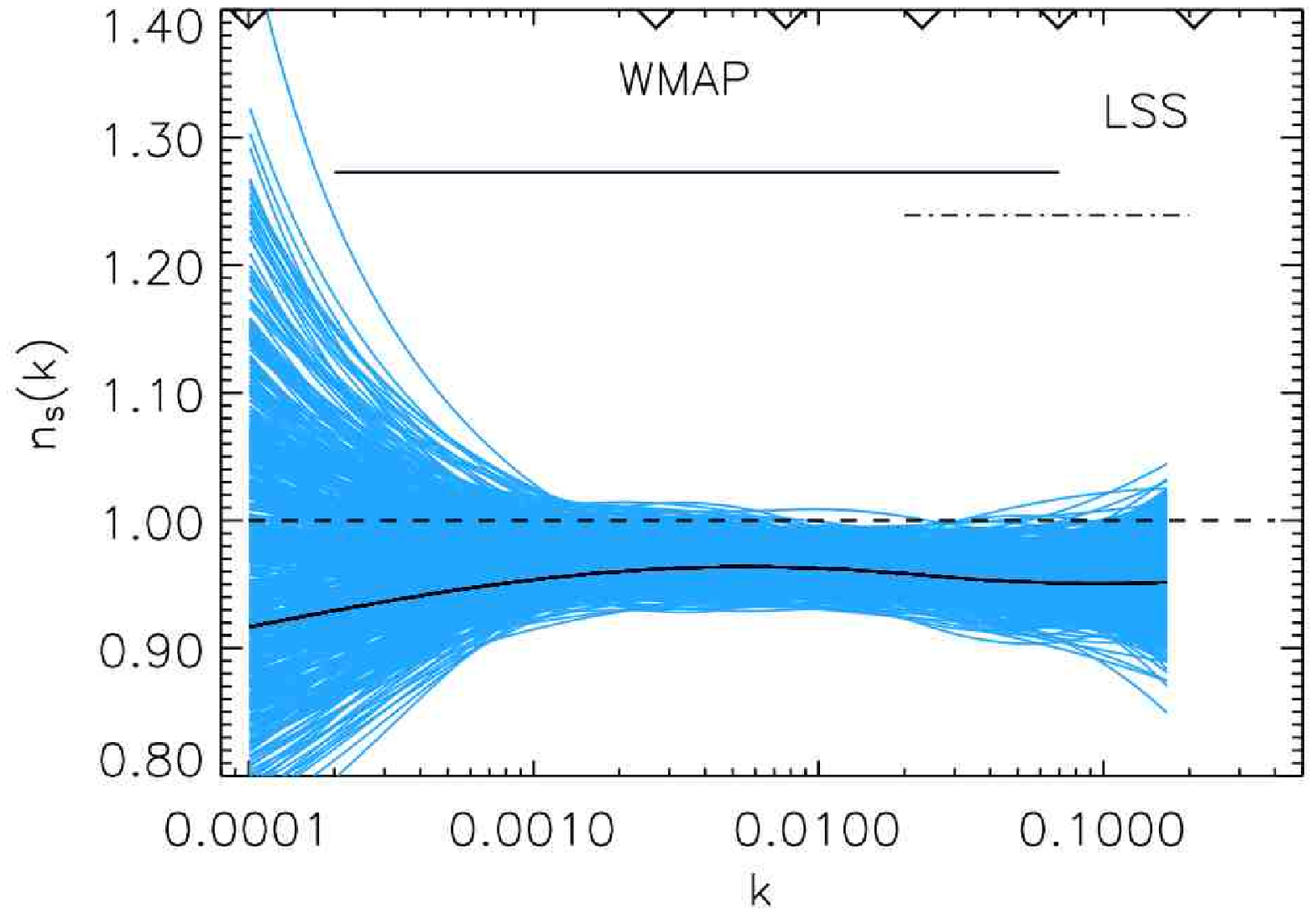}
\caption{Primordial power spectrum $P(k)$ (left) and spectral index $n_s(k)$ (right) reconstructed from WMAP3+ACBAR08 dataset  for the same penalty as the WMAPext data set. The WMAP3+ACBAR08 is very consistent with and has very similar error-bars to the full WMAPext data set. The units of $k$ are Mpc$^{-1}$.}
\label{fig:wmapacbar08}
\end{figure}

 While this work was being completed, the ACBAR collaboration released new results and the CMB temperature power spectrum for the complete set of observations \cite{ACBAR08}. These new results  greatly improve  calibration and the uncertainties on band-powers decrease by more than a factor $\sim$ 2. We show here the recovered power spectrum for the combination of WMAP3+ ACBAR08 data. We  consider only ACBAR bandpowers  which include  $ 550< \ell< 2100$. The lower $\ell$ cut is motivated by minimization of covariance with WMAP3 while the high $\ell$ cut is motivated by the ``excess" power which has been  attributed to secondary/foreground effects. We use the same penalty as for the other WMAPext runs. We find that $\Omega_b h^2$, $\Omega_c h^2$ and $h$ determinations are  virtually indistinguishable from those reported in Table \ref{tab:wmapext}, and that the reconstructed $P(k)$ and $n_s(k)$ are also virtually indistinguishable than those obtained from the full WMAPext data set (Fig. \ref{fig:wmapacbar08}). We find 
 $\sigma_8=0.79 \pm 0.04$. Note that ACBAR08 has a statistical power comparable to the entire set of other high-resolution CMB experiments used in WMAPext.
 
 \subsection{Including Large Scale Structure Data}

We implement the $n/2$-fold CV on the SDSS power spectrum band powers by taking every other  bandpower. CV shows that  while for small penalty  the CV score improves as  penalty is  increased, the  improvement flattens out at high penalties. Conservatively, we use the minimum penalty that gives the flattened-out CV score. This also happens to be the optimal penalty for  the CMBext datasets ($\lambda_{\rm opt, ext}$).

\begin{figure}
\includegraphics[width=0.53\textwidth]{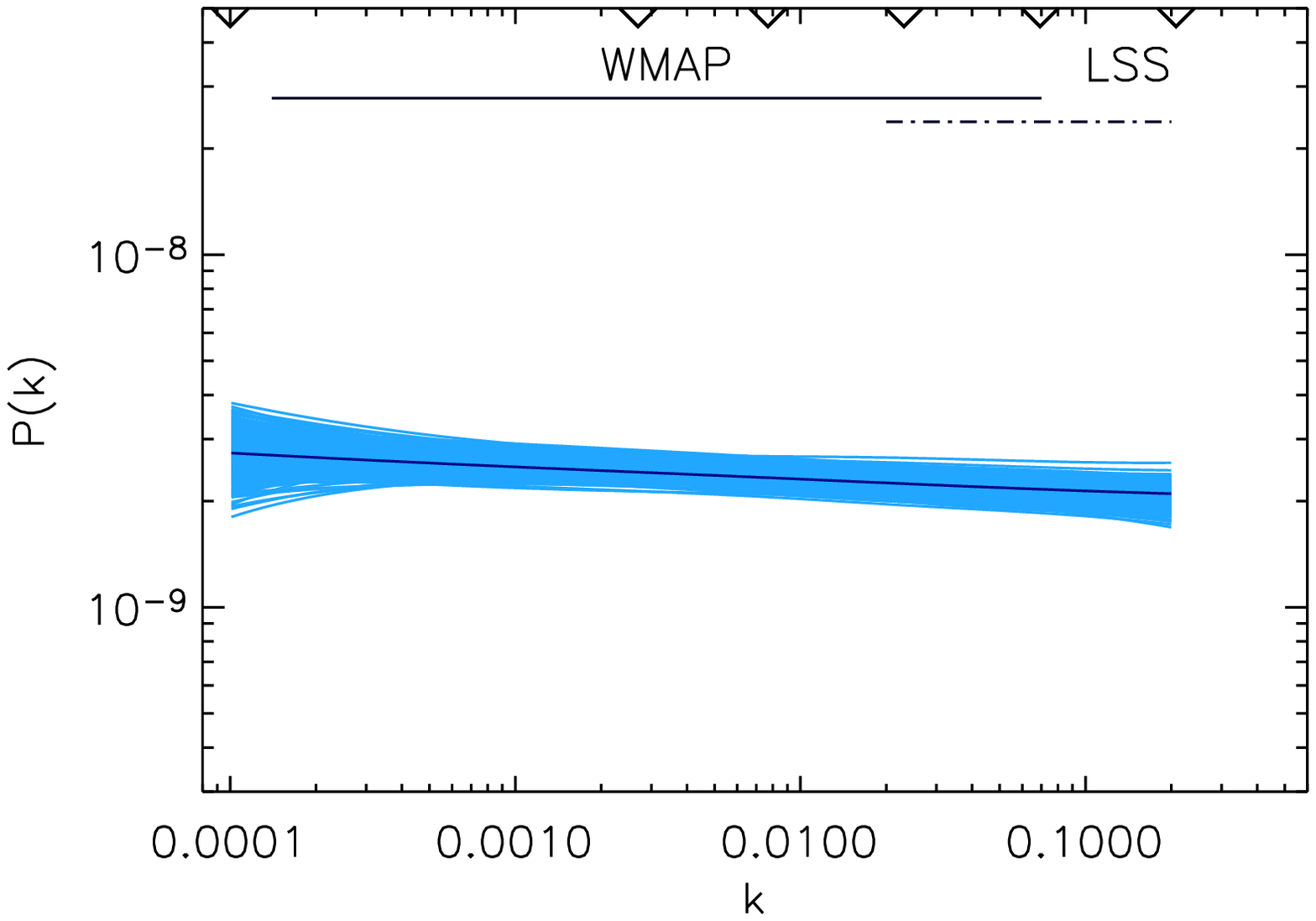}
\includegraphics[width=0.53\textwidth]{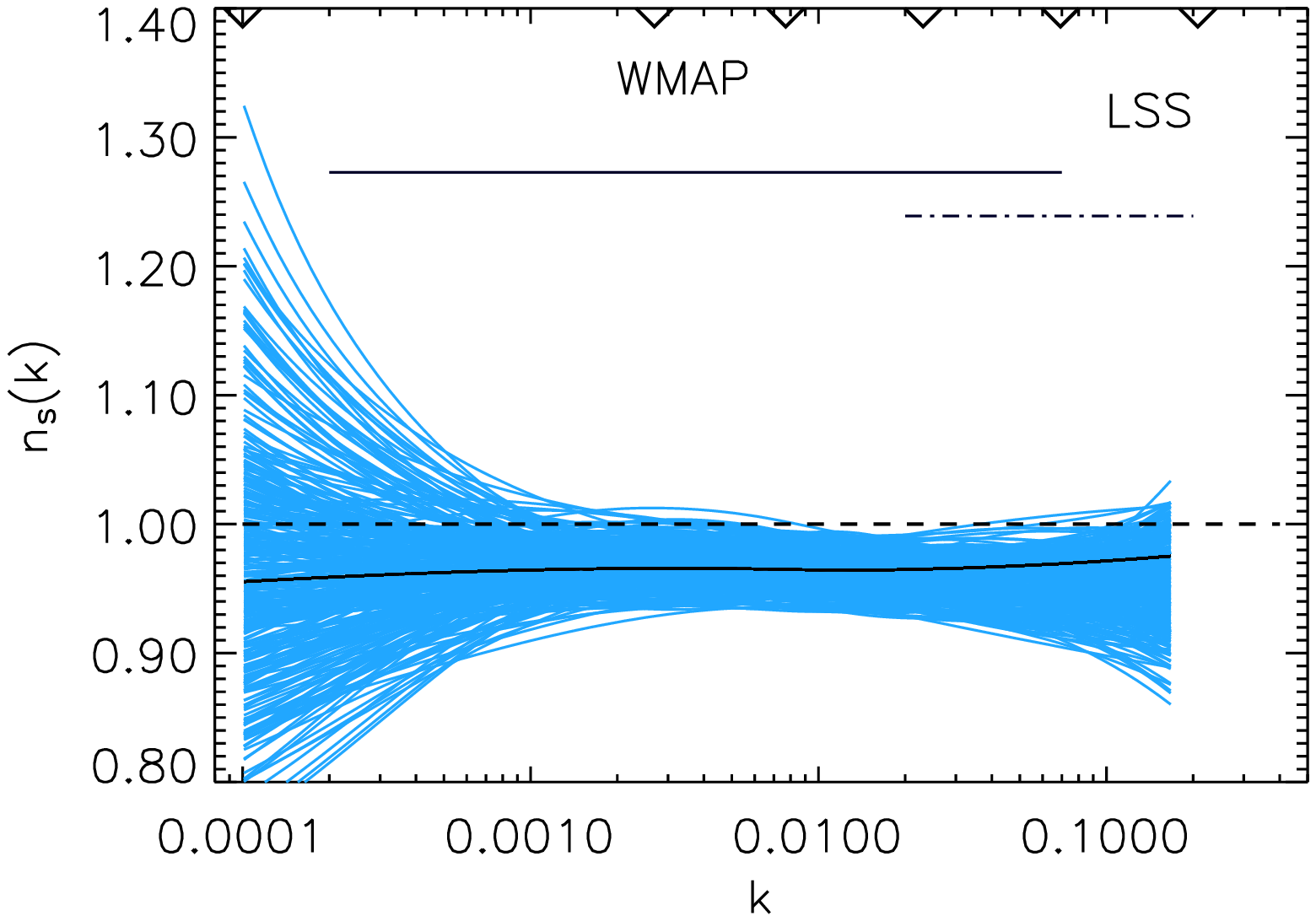}
\includegraphics[width=0.53\textwidth]{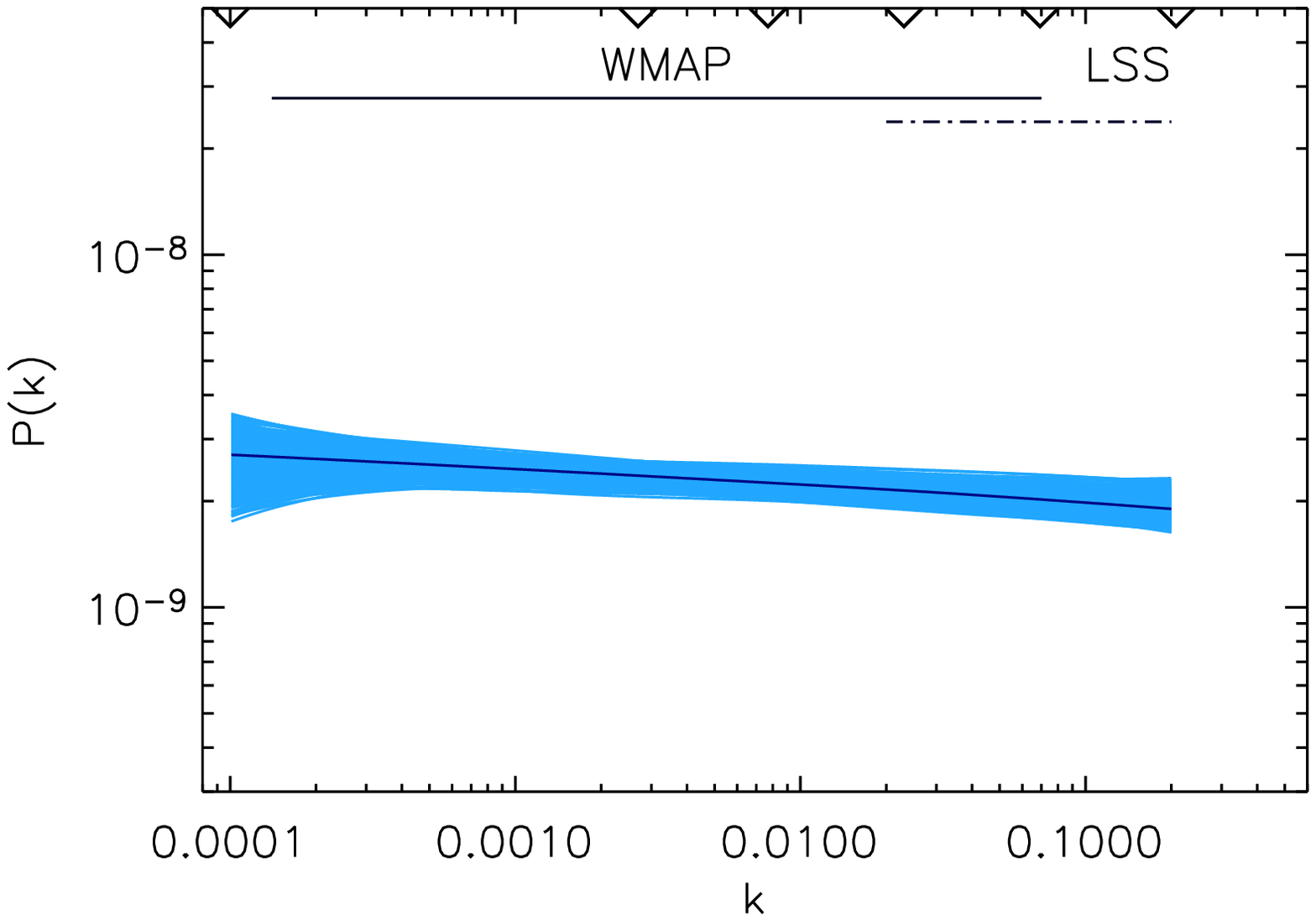}
\includegraphics[width=0.53\textwidth]{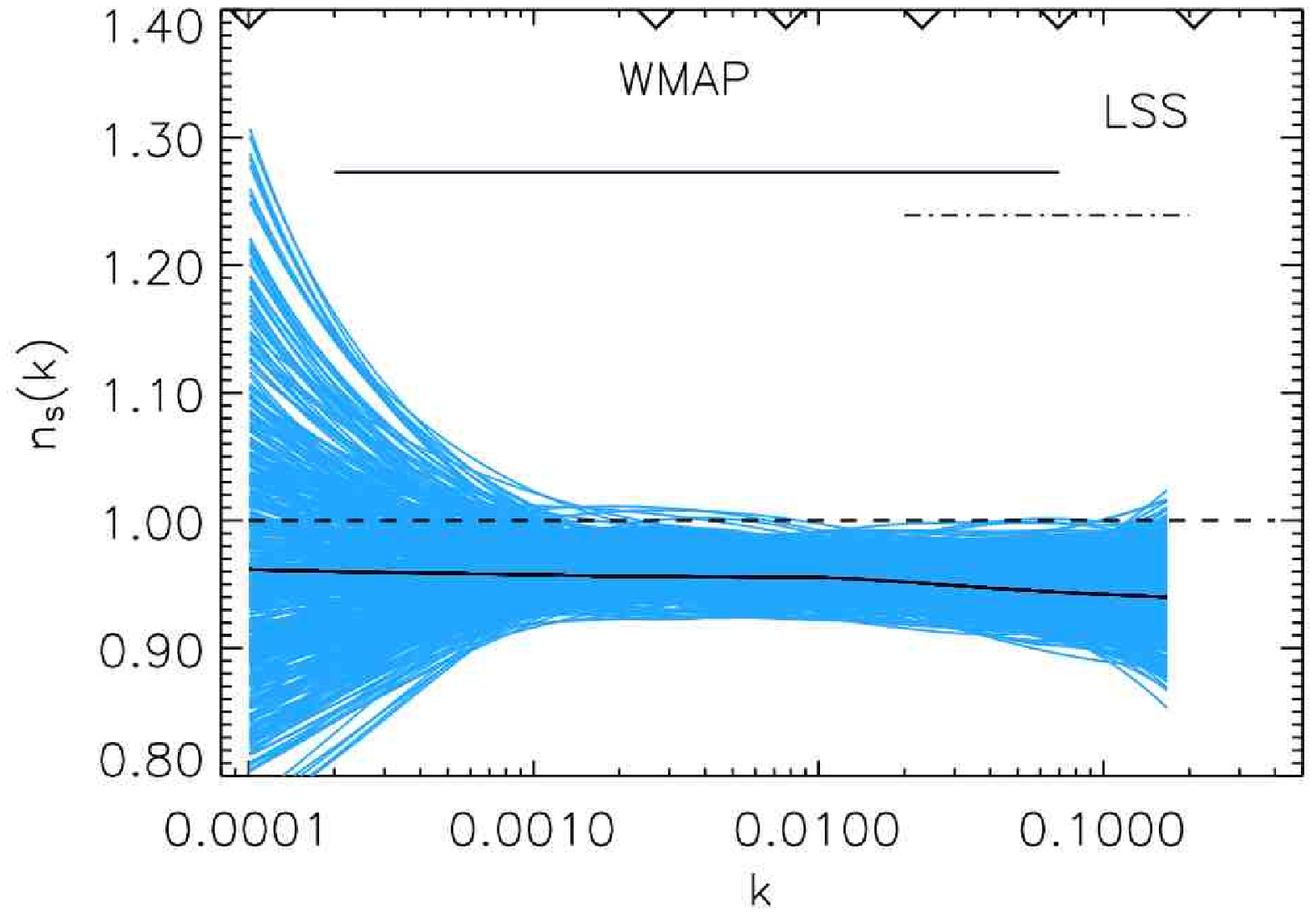}
\caption{Reconstructed power spectrum  $P(k)$ (left) and its spectral index $n_s(k)$ (right) for the WMAP3+SDSS  data set (upper panels) and WMAP3+2dFGRS data set (lower panels).}
\label{fig:pkshi5wmaplss}
\end{figure}

\begin{figure}
\includegraphics[width=0.53\textwidth]{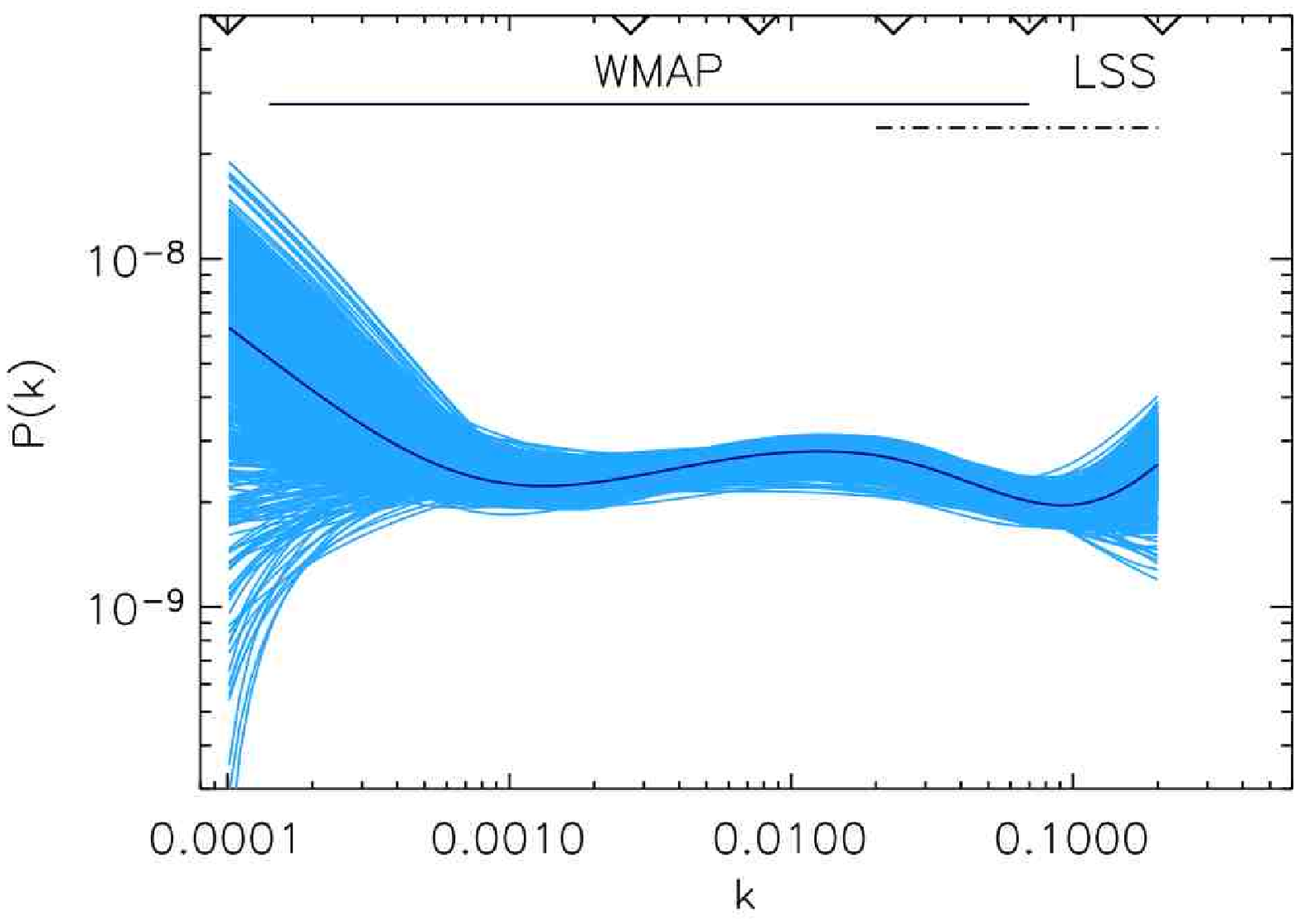}
\includegraphics[width=0.53\textwidth]{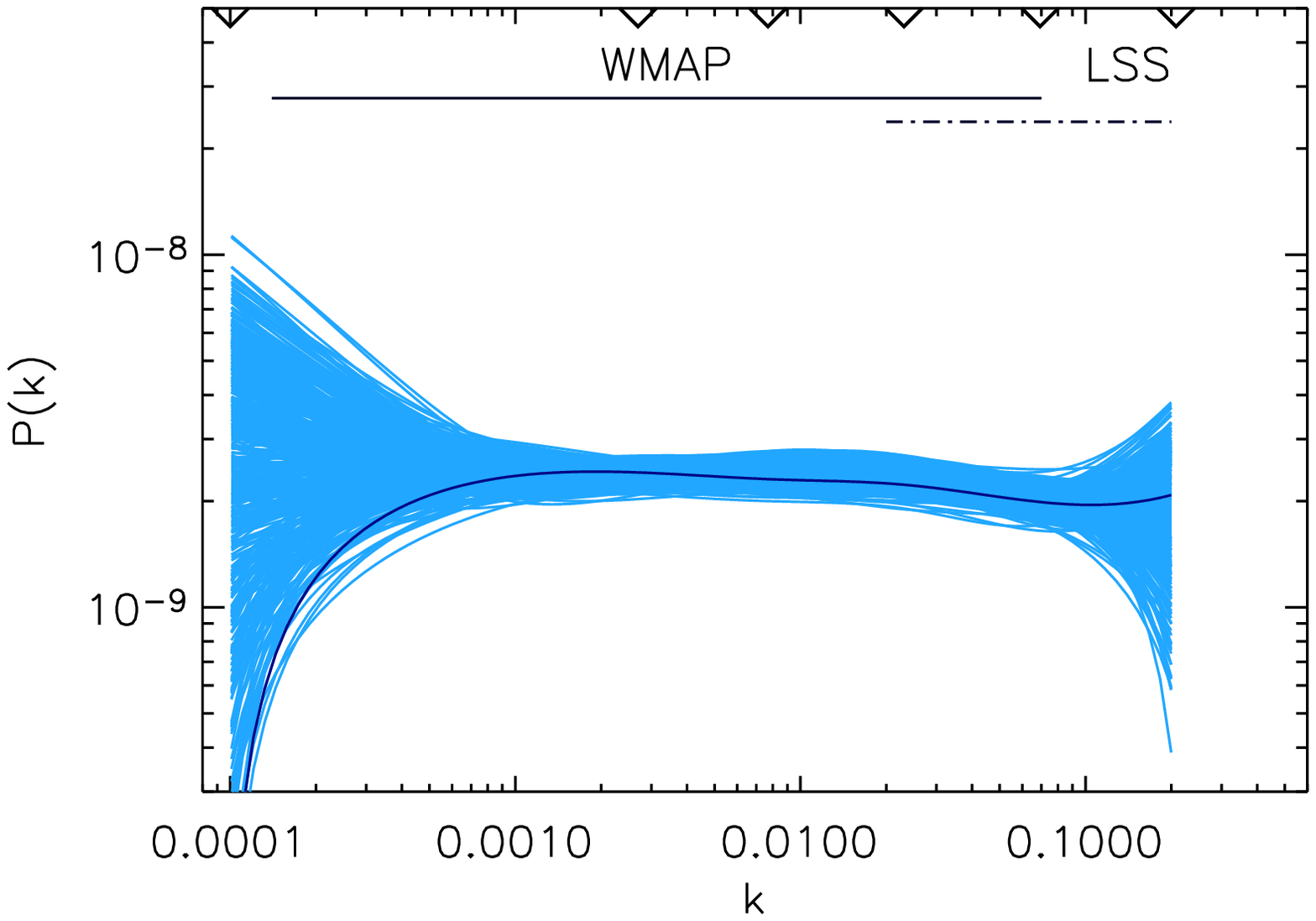}
\caption{Reconstructed power spectrum $P(k)$ with no penalty  for the WMAP3+ SDSS (left) and WMAP3+ 2dFGRS (right) data set.}
\label{fig:pklssnop}
\end{figure}

The reconstructed power spectrum from SDSS main and 2dFGRS are shown in Fig. \ref{fig:pkshi5wmaplss} (top and bottom panels, respectively). For comparison we report the reconstructed $P(k)$ without penalty in Fig. \ref{fig:pklssnop}.

The cosmological parameter constraints from  WMAP3 + large-scale structure are reported  in Table \ref{tab:wmaplss}. This dataset combination shows the same trends as  the WMAPext data combination.
 \begin{table}
\caption{\label{tab:wmaplss} Effect on cosmological parameters of the extra freedom in the primordial power spectrum  for WMAP3+SDSS main galaxy sample and WMAP3+2dFGRS data, in the same format as Table \ref{tab:wmaponly}.}
\begin{center}
\footnotesize\rm
\begin{tabular*}{\textwidth}{@{\extracolsep{\fill}}|l|c|c|c|c|}
\hline
SDSS & PL & run & spline $\lambda_{\rm opt, ext}$ & spline $\lambda=0$\\
\hline
$\Omega_b h^2$ & $0.0223\pm 0.00070$ & $0.021\pm 0.001$ & $0.0223\pm 0.00072$ & $0.0184\pm 0.0011$\\
$\Omega_c h^2$ & $0.132 \pm 0.0065$ & $0.139\pm 0.0078$ & $0.125\pm 0.0057$ & $0.14 \pm 0.012$\\
$h$  &  $0.709 \pm 0.026$ & $0.66\pm 0.03$ & $0.664 \pm 0.023$ & $0.57 \pm 0.04$\\
$\sigma_8$ & $0.77 2\pm 0.041$ & $0.783 \pm 0.041$ &  $0.86 \pm 0.037$ & $0.912 \pm 0.041$\\
\hline

\hline
2dFGRS & PL & run & spline $\lambda_{\rm opt, ext}$ & spline $\lambda=0$\\
\hline
$\Omega_b h^2$ & $0.0222\pm 0.00070$ & $0.021\pm 0.001$ &$ 0.022\pm 0.00074$&  $0.0203\pm 0.0014$  \\
$\Omega_c h^2$ & $0.126 \pm 0.0051$ & $0.128\pm 0.0055$ & $0.107\pm0.0050$ & $0.115\pm 0.011$\\
$h$  &  $0.732 \pm 0.021$ & $0.703\pm 0.026$ & $0.720\pm 0.022 $ & $0.672 \pm 0.044$ \\
$\sigma_8$ & $0.736 \pm 0.036$ & $0.739 \pm 0.038$ &$0.776\pm 0.037$&  $0.803 \pm 0.063$ \\
\hline

\end{tabular*}
\end{center}
\end{table}

Recently, a lot of attention has been given to the value of $\sigma_8$: several cosmological observables depend very strongly on this parameter, such as the number density of clusters of galaxies and  the amplitude of the contribution of the Sunyaev-Zel'dovich effect to the CMB power spectrum at mm wavelengths. Tables \ref{tab:wmaponly}, \ref{tab:wmapext} and \ref{tab:wmaplss}  show that the $\sigma_8$ determination from WMAP3 data alone depends very strongly on the assumptions about the primordial power spectrum. This can be understood if we  consider that most of the scales contributing to fluctuations on $8 h^{-1}$ Mpc are not directly probed by WMAP: an extrapolation is required.  These scales are probed by the higher resolution CMB experiments  and by large-scale structure data; thus $\sigma_8$ becomes progressively less sensitive to assumptions about the power spectrum shape.

\subsection{SDSS Luminous Red Galaxies}
Beyond the  power spectrum for the main galaxy sample,  SDSS also offers the power spectrum of the luminous red galaxies (LRGs). LRGs are more luminous than the main sample and thus probe a larger volume: the LRG power spectrum thus has potentially greater statistical power. It is, in addition, a very interesting sample to examine because many forthcoming and planned dark-energy experiments focus on these galaxies to sample even larger survey volumes and  measure the baryon acoustic oscillation (BAO) signal. \cite{ColeSanchezWilkins06, SanchezCole07} found a tension between the power spectra from the 2dFGRS sample and the LRG sample, and concluded that LRGs have a stronger scale-dependent bias than blue-selected 2dFGRS galaxies. 

\begin{figure}
\includegraphics[width=0.53\textwidth]{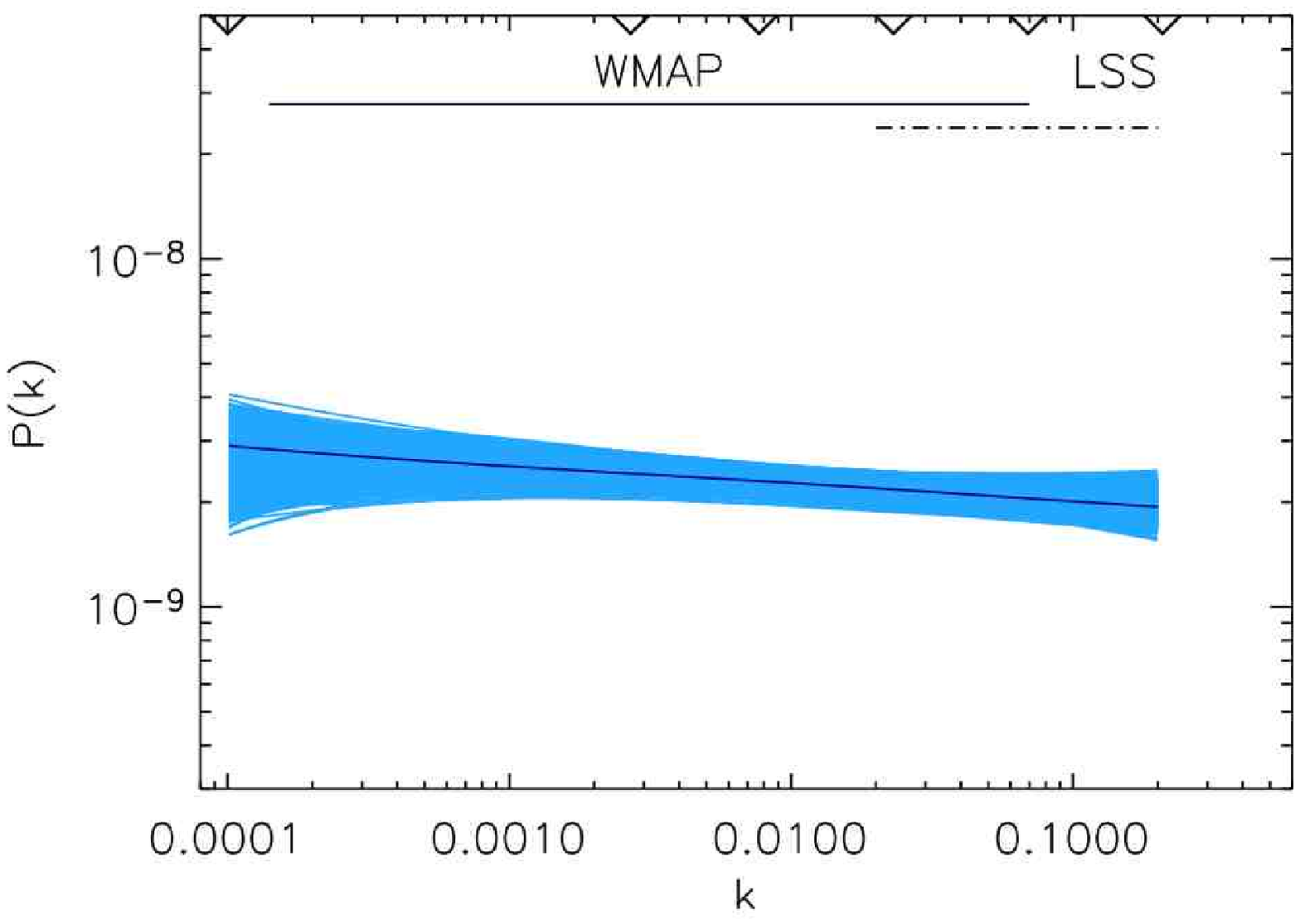}
\includegraphics[width=0.53\textwidth]{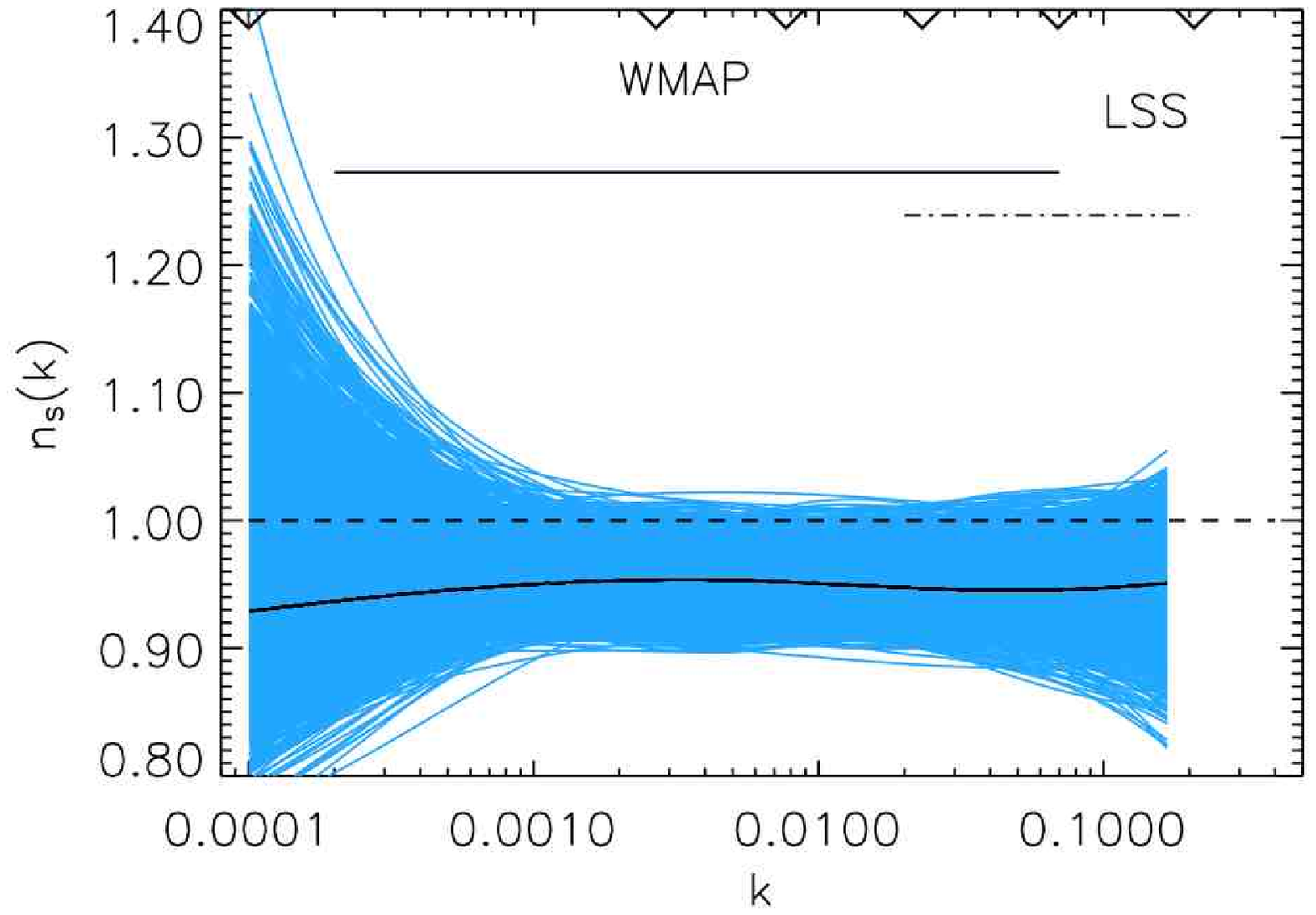}
\caption{Reconstructed power spectrum (left) and   spectral slope $n_s(k)$ (right)  for  WMAP3+ LRG for the same penalty used for the other LSS data. Comparison with Fig.\ref{fig:pkshi5wmaplss} shows that the LRG sample has less statistical power than the other LSS datasets.}
\label{fig:pklrg}
\end{figure}

Therefore, we consider the LRG sample separately and explore the recovered primordial power spectrum shape. While a full treatment and comparison between the DR4 and the DR5 power spectra will be presented in a forthcoming paper, we present here a few insights that can be enabled by a non-parametric method.
 To model  the effects of redshift space distortions, non-linearities and galaxy biasing, we use the empirical form developed in \cite{2df05}:
 \begin{equation}
 P_{\rm gal}(k)=b^2\frac{1+Qk^2}{1+A k}P_{\rm lin}(k)
 \end{equation}
 where $b$ denotes a constant scale independent normalization (bias) and $A$ and $Q$ are empirical parameters.  \cite{2df05} shows that the value $A=1.4$ is robust but that $Q$ depends on galaxy type. Thus we leave $Q$ as a free parameter; in particular we do not marginalize over it analytically with a given prior but treat it as an extra MCMC parameter. The bias parameter $b$ is treated in the same way. To work in the linear regime we consider only scales $k\le 0.1$ $h$/Mpc, and we use the same penalty as for the other LSS datasets.
We find that  the parameter $Q$ is virtually unconstrained, and that  the addition of LRG data does not improve constraints on $P(k)$ and $n_s(k)$ as much as the main SDSS sample or 2dFGRS data (Fig. \ref{fig:pklrg}).  

\begin{table}
\caption{\label{tab:lrgs} Effect on cosmological parameters of the extra freedom in the primordial power spectrum  for WMAP3+LRG. The first two columns are from \cite{TegmarkLRGDR4}.}
\begin{center}
\footnotesize\rm
\begin{tabular*}{\textwidth}{@{\extracolsep{\fill}}|l|c|c|c|c|}
\hline
WMAP+LRG & PL & run & spline $\lambda_{\rm opt, ext}$ & WMAP only, $\lambda_{\rm opt, ext}$\\
\hline
$\Omega_b h^2$ & $0.0222\pm 0.00070$ & $0.021\pm 0.001$ & $0.0225\pm 0.001$ & $0.0214\pm 0.0059$\\
$\Omega_c h^2$ & $0.105 \pm 0.004$ & $0.109\pm 0.004$ & $0.114\pm 0.006$ & $0.107\pm 0.008$\\
$h$  &  $0.73 \pm 0.019$ & $0.713\pm 0.022$ & $0.68 \pm 0.06$ & $0.72\pm 0.03$\\
$\sigma_8$ & $0.756 \pm 0.035$ & $0.739 \pm 0.036$ &  $0.80 \pm 0.04$ & $0.77 \pm 0.036$\\
\hline
\end{tabular*}
\end{center}
\end{table}

This is different to the LCDM case: when the shape of the power spectrum is fixed the statistical power of the LRG sample is greater than that of the main sample (see Tab. \ref{tab:lrgs}  and \cite{TegmarkLRGDR4, PercivalLRG}).  We  therefore conclude that a better understanding of the way LRG galaxies trace the underlying dark matter distribution is crucial  to take advantage of  the full statistical power of these data. This will be further explored elsewhere (Peiris et al. in preparation).

\section{Conclusions}

The latest compilation of cosmological data (e.g., \cite{SpergelWMAP06}) seems to indicate a significant deviation from scale invariance of the primordial power  spectrum when parameterized by a power law or by a spectral index  with a ``running". This deviation serves as a powerful tool to discriminate among theories for the origin of primordial perturbations, such as inflationary models. Primordial power spectra described by more complex functional forms have also been considered in the literature as described in \S~\ref{sec:intro}, ranging from a scale-dependence of the spectral slope (``running'') to sharp or oscillatory features (``glitches"). In interpreting the results of such studies, it is very important to have a robust criterion which allows one to determine the optimal {\sl smoothness prior} to apply to the reconstruction technique being used to describe the primordial power spectrum. Ideally, in order to minimize model-dependence, this criterion should use information from the data themselves to determine the number of degrees of freedom needed to recover the signal without fitting the noise. 

Here we build on the work of \cite{Sealfon05} and use a minimally-parametric reconstruction of the primordial power spectrum using the {\sl cross-validation} technique as the smoothness criterion. We consider a range of cosmological data -- WMAP 3-year data, complementary data from higher resolution CMB experiments: BOOMERanG, ACBAR (including the  2008 data), CBI, VSA, and large-scale structure power spectra from 2dFGRS and SDSS (both the main and LRG samples).

When considering WMAP 3-year  data alone we find indications, in agreement with \cite{Huffenbergeretal07, ACBAR08}, that the reconstructed power spectrum loses power at $k>0.02$ Mpc$^{-1}$ compared with a power law spectrum. When combining WMAP3 with either higher-resolution CMB experiments or large-scale structure data, we find no evidence from a deviation from a power law. In fact, the recovered power spectrum gives a spectral slope that is scale independent and is characterized by a red tilt, $n_s \sim 0.96$. 

As with all non-parametric methods, this approach, which does not  rely on  parameter estimation,  cannot  be used  to assess the statistical significance of a detection of deviation from scale invariance in a straightforward way. Instead, it allows one to test the sensitivity of the detection to the parametric form chosen to describe the deviation. In this context, we can interpret our findings as follows.
 In all dataset combinations WMAPext and WMAP3+LSS, the best 68\% of the spline curves are below or just touch the $n_s=1$ line over $4$ or $5$ knots  depending on the data set; this range corresponds to two or more decades in $k$. Thus naively, assuming one were to connect the knots,  one would say that the evidence is $\sim 2-2.5$ $\sigma$.   However the curves could be more ``wiggly" than simply linearly interpolating between the knots, if the data required it.  Cross-validation shows that the data do not require extra freedom in the primordial power spectrum; in addition it shows that   the data require a negligible second derivative of  $P(k)$ (i.e.  a power law $P(k)$). We should interpret this result as confirmation that  a power law power spectrum is the correct description of the data, offering renewed confidence in the $n_s$ constraints obtained by such parametric analysis.

While the spline  reconstruction used here is best suited for smooth features in the primordial power spectrum, sharp features can also be recovered if they have high enough signal-to-noise as illustrated in \cite{Sealfon05}, where a sharp step in the primordial power spectrum  was shown to be reconstructed.   However, the technique implemented here would miss features with a characteristic scale much smaller than the knot spacing unless they were highly statistically significant.

When adding either higher-resolution CMB data or LSS data to WMAP3, we find no evidence for  deviations (sharp or smooth) from a power law power spectrum.  Two independent groups \cite{Covietal06,ShafielooSouradeep07} have found   persistent features in the primordial power spectrum, but see \cite{Hamann07}.  We suggest that in general,  CV techniques could be useful to assess the statistical significance of these features.  In fact, when not using a penalty in our reconstruction, we also find ``features" in the power spectrum; these, however, go away when using  the  CV-selected penalty.

We find that, with the current data compilation, the cosmological parameters are insensitive to the extra freedom allowed here in the shape of the primordial power spectrum, with one exception: $\sigma_8$. The determination of $\sigma_8$ from WMAP3 alone is significantly affected by assumptions about the primordial power spectrum shape; while this sensitivity decreases when adding external datasets which probe smaller scales, different data combinations  lead to different results for the mean value of this parameter.

 \section*{Acknowledgments}
 LV is supported by  FP7-PEOPLE-2007-4-3-IRGn 202182 and by CSIC I3 grant 200750I034. HVP is supported by NASA through Hubble Fellowship grant \#HF-01177.01-A from the Space Telescope Science Institute, which is operated by the Association of Universities for Research in Astronomy, Inc., for NASA, under contract NAS 5-26555, by Marie Curie MIRG-CT-2007-203314, and by a STFC Advanced Fellowship. This work was partially supported by the National Center for Supercomputing Applications (NCSA) under grant number TG\verb1_1AST070005N and   utilized computational resources   on the TeraGrid (Cobalt). We acknowledge use of the Legacy Archive for Microwave Background Data (LAMBDA). Support for LAMBDA   is provided by  the NASA Office of Space Science.

 \section*{Appendix}
 
\begin{figure}
\includegraphics[width=0.53\textwidth]{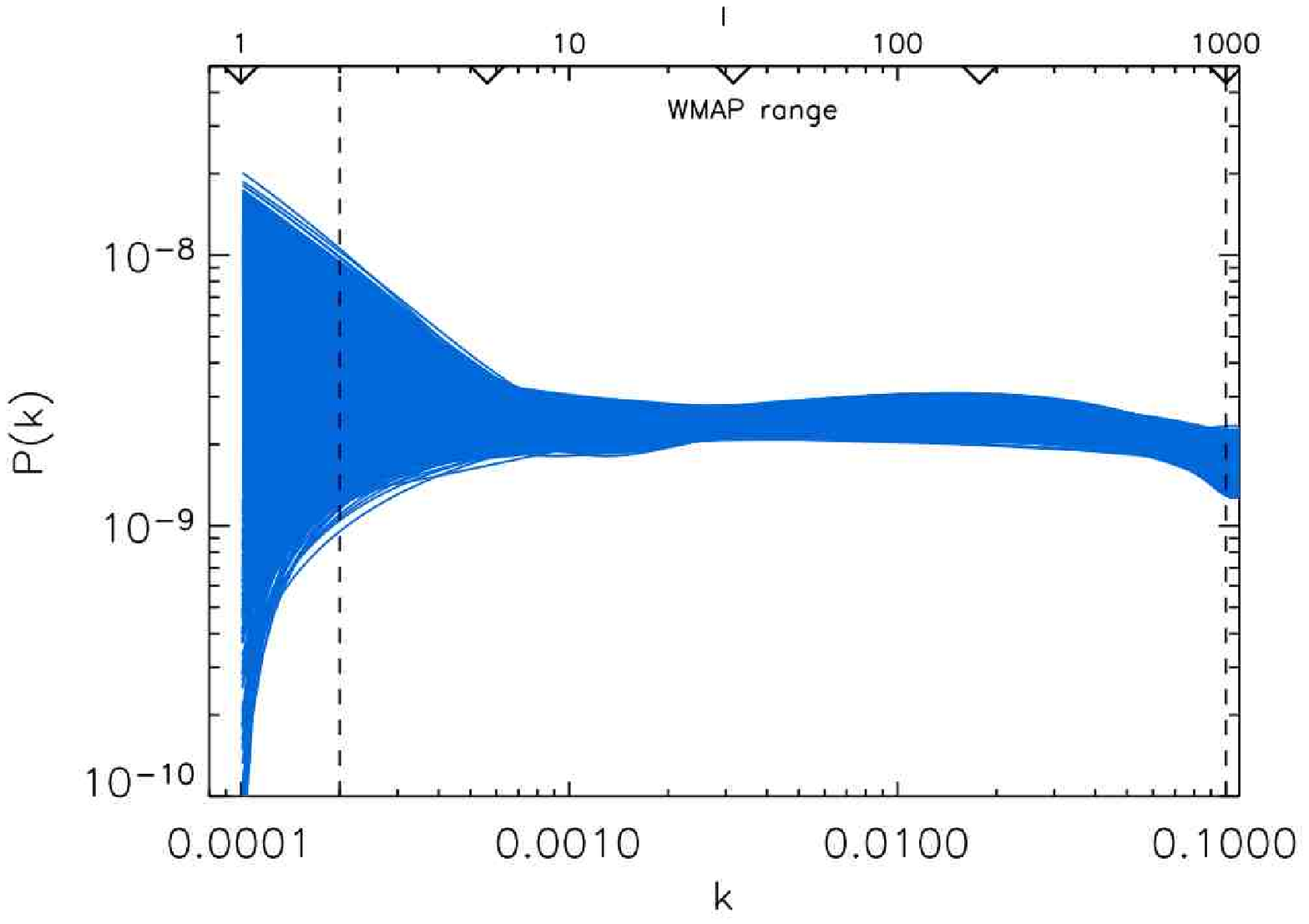}
\includegraphics[width=0.53\textwidth]{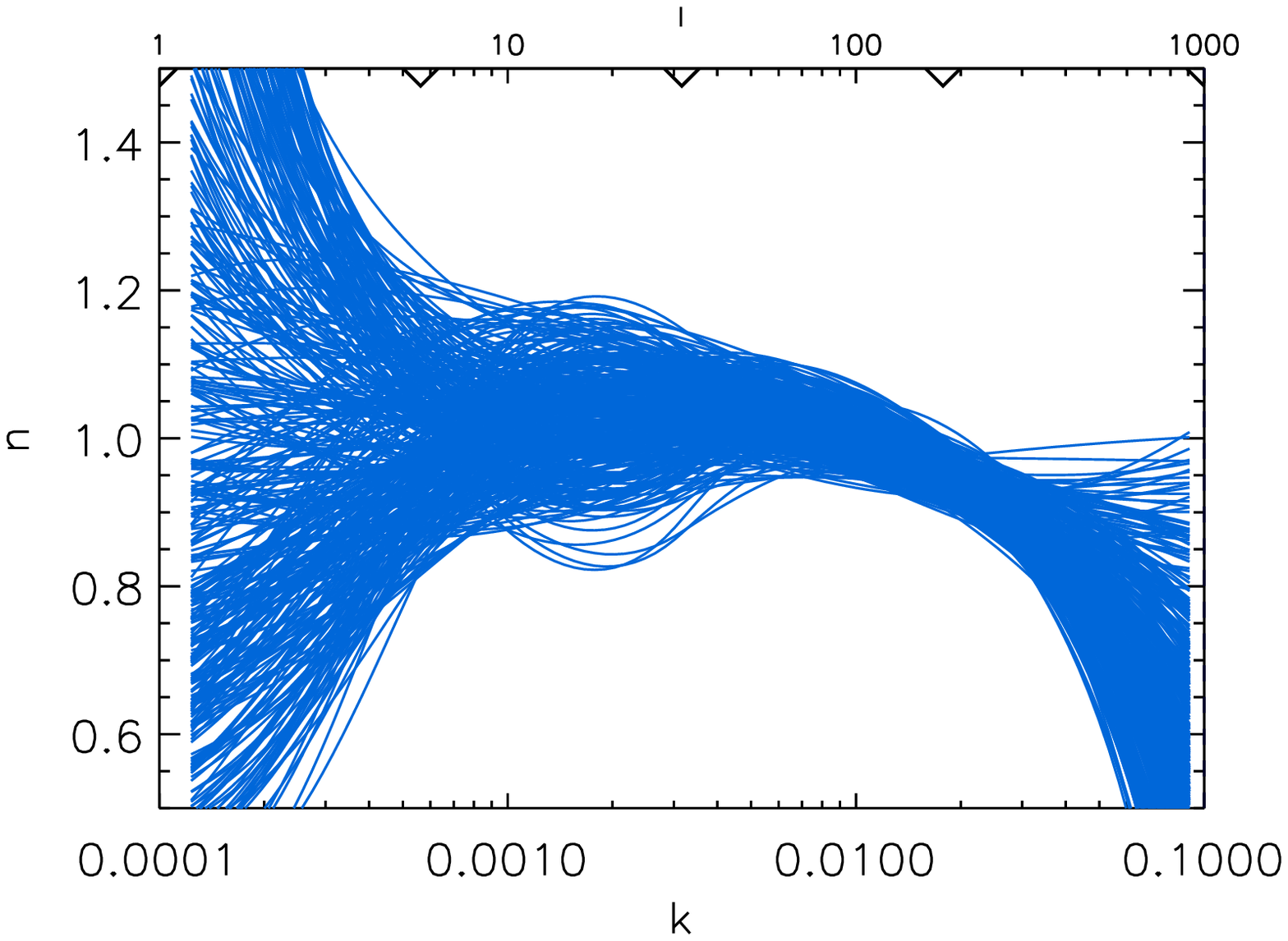}
\caption{Reconstructed power spectrum  $P(k)$ (left) and its spectral index $n_s(k)$ (right) for the WMAP3  data set using knots equally spaced in $\log k$.}
\label{fig:refereefig}
\end{figure}

 To demonstrate the insensitivity of the results to the placement of  
the knots, in Fig. \ref{fig:refereefig} we show as an example the reconstruction for  
knots equally spaced in $\log k$ for WMAP3 only data, which should be  
compared with Fig.  \ref{fig:wmappkoptpenalty}. Note that while the reconstruction is
robust to the choice of the knot positions (as long as the knots fully  
sample the $k$ range), the speed of convergence of the MCMC does depend  
on their placement.
 
 \section*{References}
 \bibliographystyle{JHEP}
 \providecommand{\href}[2]{#2}\begingroup\raggedright\endgroup

 
 \end{document}